\documentclass[conf]{new-aiaa}
\usepackage[utf8]{inputenc}
\usepackage{subdepth}
\usepackage{graphicx}
\usepackage{verbatim}
\usepackage{amsmath}
\usepackage[version=4]{mhchem}
\usepackage{siunitx}
\usepackage{longtable,tabularx}
\usepackage{multicol}
\usepackage{subcaption} 
\usepackage{algorithm}
\usepackage{algpseudocode}
\usepackage{todonotes}
\usepackage{pythonhighlight}

\usepackage{tcolorbox}
\definecolor{ucsd_blue}{RGB}{24, 43, 73}
\definecolor{ucsd_gold}{RGB}{198, 146, 20}
\definecolor{ucsd_blue_light}{RGB}{0, 98, 155}
\definecolor{ucsd_gold_light}{RGB}{255, 205, 0}
\definecolor{ucsd_gray}{RGB}{116, 118, 120}

\title{A gradient-enhanced univariate dimension reduction method for uncertainty propagation}

\author{Bingran Wang, Nicholas C. Orndorff, Mark Sperry, and John T. Hwang}
\affil{University of California, San Diego, La Jolla, CA, USA}

\begin{document}
{\begin{tcolorbox}[boxrule=0.75pt, arc=2pt, coltext=ucsd_blue, colback=white,colframe=ucsd_gold, fontupper=\rmfamily]
        \footnotesize
        \small
        \setstretch{1.25}
        {\textcolor{ucsd_gray}
        {This is the preprint version of the following article:}}
        
        Bingran Wang, Nicholas C. Orndorff, Mark Sperry, and John T. Hwang. A gradient-enhanced univariate dimension reduction method for uncertainty propagation, Aerospace Science and Technology, vol. 155, 2024, 109602.

        \vspace{1mm}
        \urlstyle{rm}
        \textcolor{ucsd_gray}{Published article:} \;
        \url{https://doi.org/10.1016/j.ast.2024.109602}

        \textcolor{ucsd_gray}
        {Preprint pdf:} \; 
        \url{https://github.com/LSDOlab/lsdo_bib/blob/main/pdf/wang2024gudr.pdf}


        \textcolor{ucsd_gray}
        {Bibtex:} \;    \url{https://github.com/LSDOlab/lsdo_bib/blob/main/bib/wang2024gudr.bib}

    \end{tcolorbox}
    
    \vspace{-6mm}
}

\maketitle
 \begin{abstract}
The univariate dimension reduction (UDR) method stands as a way to estimate the statistical moments of the output that is effective in a large class of uncertainty quantification (UQ) problems. 
UDR's fundamental strategy is to approximate the original function using univariate functions so that the UQ cost only scales linearly with the dimension of the problem. 
Nonetheless, UDR's effectiveness can diminish when uncertain inputs have high variance, particularly when assessing the output's second and higher-order statistical moments.
This paper proposes a new method, gradient-enhanced univariate dimension reduction (GUDR), that enhances the accuracy of UDR by incorporating univariate gradient function terms into the UDR approximation function. 
Theoretical results indicate that the GUDR approximation is expected to be one order more accurate than UDR in approximating the original function, and it is expected to generate more accurate results in computing the output's second and higher-order statistical moments.
Our proposed method uses a computational graph transformation strategy to efficiently evaluate the GUDR approximation function on tensor-grid quadrature inputs, and use the tensor-grid input-output data to compute the statistical moments of the output.
With an efficient automatic differentiation method to compute the gradients, our method preserves UDR's linear scaling of computation time with problem dimension.
Numerical results show that the GUDR is more accurate than UDR in estimating the standard deviation of the output and has a performance comparable to the method of moments using a third-order Taylor series expansion.
\end{abstract}
\section{Introduction}

Uncertainties are inherent in numerous scientific and engineering problems.
Spanning from fields such as weather forecasting~\cite{joslyn2010communicating,hess-9-381-2005}, machine learning~\cite{hullermeier2021aleatoric}, structural analysis~\cite{wan2014analytical,hu2018uncertainty} and aircraft design~\cite{ng2016monte,wang2024graph,lim2022uncertainty}, these uncertainties can have a considerable effect on a system's behavior or performance.
Forward uncertainty quantification (UQ), or uncertainty propagation, aims to evaluate how uncertainties in inputs influence the system's outputs or quantities of interest (QoIs).
The input uncertainties can arise either from variations in operating conditions and model parameters, known as aleatoric uncertainties, or from model errors attributable to incomplete knowledge, known as epistemic uncertainties.
By quantifying the impact of the input uncertainties, UQ provides critical support for decision-making and risk assessment in designing engineering systems.
In this paper, we consider a UQ  problem within a probabilistic framework, with the input uncertainties characterized using continuous probability density distributions. 
The objective here is to estimate the QoIs' statistical moments.

Developing computational methods to solve UQ problems has been a popular research topic for decades, and many approaches have been proposed to tackle UQ problems in different scenarios.
The most common UQ methods include the method of moments, polynomial chaos, kriging, and Monte Carlo.
The method of moments~\cite{wooldridge2001applications,fragkos2019pfosm} uses Taylor series approximations to compute statistical moments cost-effectively, but it may lack accuracy for large input variances and becomes impractical at higher expansion orders due to increased computational demand.
Monte Carlo methods are commonly used for solving high-dimensional UQ problems as their convergence rate is unaffected by the dimensionality of the input space.
Recent advancements seek to enhance its efficiency through multi-fidelity~\cite{peherstorfer2016optimal,peherstorfer2018survey} and importance sampling techniques~\cite{tabandeh2022review}. 
However, for low-dimensional scenarios, Monte Carlo may require an excessive number of model evaluations to match the precision of other UQ methods.
Kriging, or Gaussian process regression, creates a surrogate response surface from input-output data, facilitating extensive model evaluations for reliability analysis~\cite{kaymaz2005application,hu2016single} or optimization under uncertainty~\cite{rumpfkeil2013optimizations}. 
The polynomial chaos approach models the QoI using orthogonal polynomials based on the distributions of the inputs, leveraging the smoothness in the random space for swift convergence with sampling or integration~\cite{hosder2006non,jones2013nonlinear,keshavarzzadeh2017topology}.
Both kriging and polynomial chaos are favored for low- to medium-dimensional problems but their computational cost does not scale well with the dimension of the problem, limiting their direct application in complex UQ tasks.

In tackling high-dimensional UQ problems, dimension reduction techniques offer a strategic means to curtail computational expenses. Among these, sensitivity analysis~\cite{morio2011global}, partial least squares (PLS)~\cite{chun2010sparse}, the active subspace (AS) method~\cite{constantine2014active}, and univariate dimension reduction (UDR)~\cite{rahman2004univariate} are widely used.
Sensitivity analysis, PLS, and AS leverage function or gradient evaluations at sampled input points to isolate key uncertain inputs or to uncover lower-dimensional latent variables in order to reduce the dimension of the problem. However, the success of these methods hinges on the existence of a compact set of latent variables that effectively capture the function's behavior.
In contrast, UDR approximates the function as a sum of univariate functions, transforming a multidimensional integration problem into multiple one-dimensional problems. Consequently, the computational load of UDR increases only linearly with the number of dimensions, presenting a very cost-effective method for estimating the statistical moments of the QoIs.
UDR has also been adapted for calculating complete probability distributions~\cite{piric2015reliability} and reliability metrics of QoIs~\cite{huang2006uncertainty,lee2008dimension}. 
Nevertheless, its precision diminishes with increasing input variances, particularly when estimating second-order and higher statistical moments.
To enhance UDR's accuracy, some extensions incorporate bivariate functions into the UDR approximation~\cite{xu2004generalized}. Although these adjustments provide better accuracy, they compromise UDR's linear scalability, presenting a trade-off that requires careful consideration in the method's application.

Wang et al. recently introduced a novel method known as \textit{Accelerated Model evaluations on Tensor grids using Computational graph transformations} (AMTC)~\cite{wang2023accelerating, wang2022efficient}. 
This method reduces the model evaluation cost on tensor-grid inputs by modifying the computational graph of the model.
This modification eliminates the redundant evaluations on the operation level that are incurred by the tensor structure of the inputs. 
The existing work applies AMTC to integration-based non-intrusive polynomial chaos, assuming a certain type of sparsity is present in the computational graph of the original computational model.
In some problems, this sparsity enables speedups of multiple orders of magnitude~\cite{wang2024extension, wang2024graphpartial, wang2024graphthesis}.
However, if the computational graph does not possess this type of sparsity, the AMTC-based UQ methods may be outperformed by other UQ methods.

This paper proposes a new UQ method, named gradient-enhanced univariate dimension reduction (GUDR), intending to enhance the accuracy of UDR by incorporating univariate gradient functions into the approximations. 
The GUDR approximation utilizes univariate function evaluations, univariate gradient evaluations, and a single Hessian evaluation at the mean values of the inputs. 
Theoretical results show that the GUDR approximation is one order more accurate in estimating the original function, and is expected to have a comparable level of accuracy as the third-order Taylor series expansion in estimating the second and higher-order statistical moments of the output.
To estimate the statistical moments of the output, we propose to use the AMTC strategy to effectively perform the evaluations of the GUDR approximation function on tensor-grid quadrature points.
This tensor-grid input-output data can also be used to compute the risk measures of the output.
The results show that, with an efficient automatic differentiation method, the computational cost of the GUDR method scales linearly with the dimension of the problem.

This method has been applied to {five} UQ problems: one 2D and one 3D mathematical function, one mathematical function with varying dimensions, one 4D rotor aerodynamic analysis problem, and one 7D problem derived from a practical aircraft design scenario. 
The results indicate that on the {first two} mathematical functions, when estimating the standard deviation of the output, GUDR is more accurate than UDR and has comparable performance as the method of moments method using third-order Taylor expansion.
{On the third mathematical function, numerical results show that the function evaluation cost of GUDR scales linearly with the problem dimension and is less than four times that of UDR.}
On the 4D rotor analysis and 7D aircraft design problems, GUDR enhances the accuracy of UDR by an order of magnitude in estimating the standard deviation and becomes the most cost-effective method to use among the UQ methods implemented.

This paper is organized as follows. 
Section \ref{Sec: Background} provides some background for UDR and AMTC. 
Section \ref{Sec: Methodology} presents the details of the GUDR method. 
Section \ref{Sec: Numerical Results} shows numerical results on the test problems.
Section \ref{Sec: Conclusion} summarizes the work and offers concluding thoughts.
\section{Background}
\label{Sec: Background}
\subsection{Univariate dimension reduction}
We consider an uncertainty quantification (UQ) problem involving a function $f(u)$, where $u \in \mathbb{R}^d$ denotes the input vector, and $f \in \mathbb{R}$ represents a scalar output.
The uncertain inputs are expressed as a stochastic vector denoted by $U := [U_1, \ldots, U_d]^T$, with the assumption that these random variables are mutually independent. The stochastic input vector has probability density distribution $\rho(u)$ with support $\Gamma$. 
The UQ problem aims to compute the statistical moments of the output random variable, $f(U)$.
Given that the mean of the uncertain inputs vector is $\mu:= [\mu_1, \ldots, \mu_d]^T$, the univariate dimension reduction (UDR)~\cite{rahman2004univariate} method approximates the original function $f(u)$  using a sum of univariate functions:
\begin{equation}
\label{eqn:udr}
\hat{f}(u) = \sum_{i=1}^d f_i(u_i) - (d-1) f(\mu),
\end{equation}
where each univariate function term is defined as $f_i(u_i): = f(\mu_1, \ldots, \mu_{i-1}, u_i, \mu_{i+1}, \ldots, \mu_{d})$.
Expanding $f(u)$ in a Taylor series at $u =\mu$ yields
\begin{equation}
\label{eqn:general_taylor_original}
\begin{aligned}
f(u)  & = f(\mu) + \sum_{j=1}^{\infty} \sum_{i=1}^d \frac{1}{j!} \frac{\partial^j f}{\partial u_i^j} (\mu) (u_i-\mu_i)^j
+ \sum_{j_2=1}^{\infty} \sum_{j_1=1}^{\infty} \frac{1}{j_1!j_2!} \sum_{i_1 = 1}^{d-1} \sum_{i_2 > i_1}^{d}\frac{\partial^{j_1+j_2} f}{\partial u_{i_1}^{j_1} \partial u_{i_2}^{j_2}} (\mu) (u_{i_1} -\mu_{i_1})^{j_1} (u_{i_2}-\mu_{i_2})^{j_2} \\
& + \sum_{j_3=1}^{\infty} \sum_{j_2=1}^{\infty} \sum_{j_1=1}^{\infty} \frac{1}{j_1!j_2!j_3!} \sum_{i_1 = 1}^{d-2} \sum_{i_2 > i_1}^{d-1} \sum_{i_3 > i_2}^d \frac{\partial^{j_1+j_2+j_3} f}{\partial u_{i_1}^{j_1} \partial u_{i_2}^{j_2} \partial u_{i_3}^{j_3}} (\mu) (u_{i_1}-\mu_{i_1})^{j_1} (u_{i_2}-\mu_{i_2})^{j_2} (u_{i_3}-\mu_{i_3})^{j_3} + \dots\\
\end{aligned}
\end{equation}
The Taylor series expansion of the UDR approximation function, $\hat{f}(u)$, at $u =\mu$ can be expressed as
\begin{equation}
\label{eqn:general_taylor_UDR}
\begin{aligned}
\hat{f}(u)  & = f(\mu) + \sum_{j=1}^{\infty} \sum_{i=1}^d \frac{1}{j!} \frac{\partial^j f}{\partial u_i^j} (\mu) (u_i-\mu_i)^j.
\end{aligned}
\end{equation}
If we compare the Taylor series of the original function with the UDR approximation function, 
all of the Taylor series terms in \eqref{eqn:general_taylor_UDR} are contained in \eqref{eqn:general_taylor_original}, and the residual errors are
\begin{equation}
\label{eqn: general_residual_error}
\begin{aligned}
    f(u) - \hat{f}(u) & = \sum_{j_2=1}^{\infty} \sum_{j_1=1}^{\infty} \frac{1}{j_1!j_2!} \sum_{i_1 = 1}^{d-1} \sum_{i_2 > i_1}^{d}\frac{\partial^{j_1+j_2} f}{\partial u_{i_1}^{j_1} \partial u_{i_2}^{j_2}} (\mu) (u_{i_1} -\mu_{i_1})^{j_1} (u_{i_2}-\mu_{i_2})^{j_2} \\
& + \sum_{j_3=1}^{\infty} \sum_{j_2=1}^{\infty} \sum_{j_1=1}^{\infty} \frac{1}{j_1!j_2!j_3!} \sum_{i_1 = 1}^{d-2} \sum_{i_2 > i_1}^{d-1} \sum_{i_3 > i_2}^d \frac{\partial^{j_1+j_2+j_3} f}{\partial u_{i_1}^{j_1} \partial u_{i_2}^{j_2} \partial u_{i_3}^{j_3}} (\mu) (u_{i_1}-\mu_{i_1})^{j_1} (u_{i_2}-\mu_{i_2})^{j_2} (u_{i_3}-\mu_{i_3})^{j_3} + \dots\\
\end{aligned}
\end{equation}
This shows that the UDR approximation is only second-order accurate in approximating the original function~\cite{rahman2004univariate}.
However, when it comes to estimating the mean of the output, the residual errors become
\begin{equation}
\begin{aligned}
\label{eqn:general_udrerror_mean}
    \mathbb{E}[f(U)] -  \mathbb{E}[\hat{f}(U)] & =  \int_{\Gamma_1} \ldots \int_{\Gamma_d}  \left(f(u) - \hat{f}(u) \right)\rho(u)du_1\ldots du_d \\
    & = \frac{1}{2!2!} \sum_{i_1 = 1}^{d-1} \sum_{i_2 > i_1}^{d}\frac{\partial^{4} f}{\partial u_{i_1}^{2} \partial u_{i_2}^{2}} (\mu) \mathbb{E}[(u_{i_1} -\mu_{i_1})^{2} (u_{i_2}-\mu_{i_2})^{2}] + \ldots,
\end{aligned}
\end{equation}
which involves only fourth and higher-order integration terms.
In contrast, the 3rd-order Taylor series method approximates the function as
\begin{equation}
\label{eqn:general_3d_taylor}
    \begin{aligned}
\Tilde{f}(u)  & = f(\mu) + \sum_{j=1}^{3} \sum_{i=1}^d \frac{1}{j!} \frac{\partial^j f}{\partial u_i^j} (\mu) (u_i-\mu_i)^j
+ \sum_{j_1 = 1}^2\sum_{j_2 = 1}^{j_1+j_2\leq3}\frac{1}{j_1!j_2!} \sum_{i_1 = 1}^{d-1} \sum_{i_2 > i_1}^{d}\frac{\partial^{j_1+j_2} f}{\partial u_{i_1}^{j_1} \partial u_{i_2}^{j_2}} (\mu) (u_{i_1} -\mu_{i_1})^{j_1} (u_{i_2}-\mu_{i_2})^{j_2} \\
& +  \sum_{i_1 = 1}^{d-2} \sum_{i_2 > i_1}^{d-1} \sum_{i_3 > i_2}^d \frac{\partial^{3} f}{\partial u_{i_1} \partial u_{i_2} \partial u_{i_3}} (\mu) (u_{i_1}-\mu_1) (u_{i_2}-\mu_2)(u_{i_3}-\mu_3). \\
\end{aligned}
\end{equation}
When estimating the mean of the output with this method,
the residual errors are
\begin{equation}
\begin{aligned}
\label{eqn:general_t3error_mean}
    \mathbb{E}[f(U)] -  \mathbb{E}[\Tilde{f}(U)] & = \frac{1}{2!2!} \sum_{i_1 = 1}^{d-1} \sum_{i_2 > i_1}^{d}\frac{\partial^{4} f}{\partial u_{i_1}^{2} \partial u_{i_2}^{2}} (\mu) \mathbb{E}[(u_{i_1} -\mu_{i_1})^{2} (u_{i_2}-\mu_{i_2})^{2}]   \\
    & + \sum_{i=1}^d \frac{1}{4!} \frac{\partial^4 f}{\partial u_i^4} (\mu)  \mathbb{E}[(u_i-\mu_i)^4] + \ldots
\end{aligned}
\end{equation}
Upon comparing the UDR's residual errors in \eqref{eqn:general_udrerror_mean} with those of the 3rd-order Taylor series in \eqref{eqn:general_t3error_mean}, we observe that both methods are fourth-order accurate in estimating the mean of the output. However, the UDR comprises fewer fourth-order integration terms. If we assume the constants on the fourth-order terms are similar, it suggests that UDR generally yields a more accurate result than the 3rd-order Taylor expansion when estimating the mean of the output.

Nevertheless, for higher-order statistical moments, the UDR approximation may not perform as well as the 3rd-order Taylor expansion. For instance, when estimating the second-order statistical moments, the UDR's relative errors are:
\begin{equation}
\begin{aligned}
\label{eqn:general_udrerror_sd}
    \mathbb{E}[f(U)^2] -  \mathbb{E}[\hat{f}(U)^2] & = 
    \sum_{i_1 = 1}^{d-1} \sum_{i_2>i_1}^{d} \left( \frac{\partial^2 f}{\partial u_{i_1} \partial u_{i_2}}(\mu)\right)^2 \mathbb{E}[(u_{i_1}-\mu_{i_1})^2 (u_{i_2}-\mu_{i_2})^2]
    \\
    & + \sum_{i_1 = 1}^{d-1} \sum_{i_2>i_1}^{d} \left( \frac{\partial f}{\partial u_{i_1}}(\mu) \frac{\partial^3 f}{\partial u_{i_1} \partial u_{i_2}^2}(\mu) + \frac{\partial f}{\partial u_{i_2}}(\mu) \frac{\partial^3 f}{\partial u_{i_1}^2 \partial u_{i_2}}(\mu)  \right) \mathbb{E}[(u_{i_1}-\mu_{i_1})^2 (u_{i_2}-\mu_{i_2})^2]  \\
    & + \frac{1}{2!}f(\mu) \sum_{i_1 = 1}^{d-1} \sum_{i_2 > i_1}^{d}\frac{\partial^{4} f}{\partial u_{i_1}^{2} \partial u_{i_2}^{2}} (\mu) \mathbb{E}[(u_{i_1} -\mu_{i_1})^{2} (u_{i_2}-\mu_{i_2})^{2}] +\ldots\\
\end{aligned}
\end{equation}
In comparison, the residual errors for the 3rd order Taylor series method are
\begin{equation}
\label{eqn:general_t3error_sd}
     \mathbb{E}[f(U)^2] -  \mathbb{E}[\Tilde{f}(U)^2] =  \frac{1}{2!}f(\mu) \sum_{i_1 = 1}^{d-1} \sum_{i_2 > i_1}^{d}\frac{\partial^{4} f}{\partial u_{i_1}^{2} \partial u_{i_2}^{2}} (\mu) \mathbb{E}[(u_{i_1} -\mu_{i_1})^{2} (u_{i_2}-\mu_{i_2})^{2}] +\ldots
\end{equation}
Comparing the residual errors of the 3rd-order Taylor series method in \eqref{eqn:general_t3error_sd} with the residual errors of the UDR in \eqref{eqn:general_udrerror_sd}, both of the methods are also fourth-order accurate in estimating the second-order moment of the output, but the relative errors of the UDR comprise significantly more fourth-order integration terms compared with the 3rd-order Taylor series expansion. Thus it is expected that the UDR is not as accurate as the 3rd Taylor series expansion method when used to estimate the second order moment of the output. 

The detailed derivation of the residual errors for the UDR approximation in estimating the first and second-order statistical moments for a 2D problem can be found in Appendix~\ref{Sec: UDR_2D}.

\subsubsection{Computational cost}
The major advantage of the UDR approximation is that when estimating the statistical moments of the output, the corresponding multidimensional integration problem can be decomposed into multiple one-dimensional integration problems, which significantly reduces the required computational cost.
For example, when computing the mean of the output, the UDR yields
\begin{equation}
    \mathbb{E}[\hat{f}(U)] = \sum_{i=1}^d  \int_{\Gamma_i}
    f_i(u_i)\rho(u_i)du_i - (d-1) f(\mu).
\end{equation}
This only requires performing $d$ one-dimensional integrations instead of performing a $d$-dimensional integration. If we use the quadrature rule to approximate each integral with $k$ quadrature points, the total number of model evaluations required is $ kd + 1$, which scales only linearly with the dimension of the UQ problem. For higher-order statistical moments of the output, UDR can compute it following a recurring formula in \cite{rahman2004univariate} without requiring extra model evaluations.

\subsection{Computational graph transformations to accelerate tensor-grid evaluations}

\begin{figure}%
    \centering
    \subfloat[\centering Computational graph without using AMTC]{{\includegraphics[width=5cm]{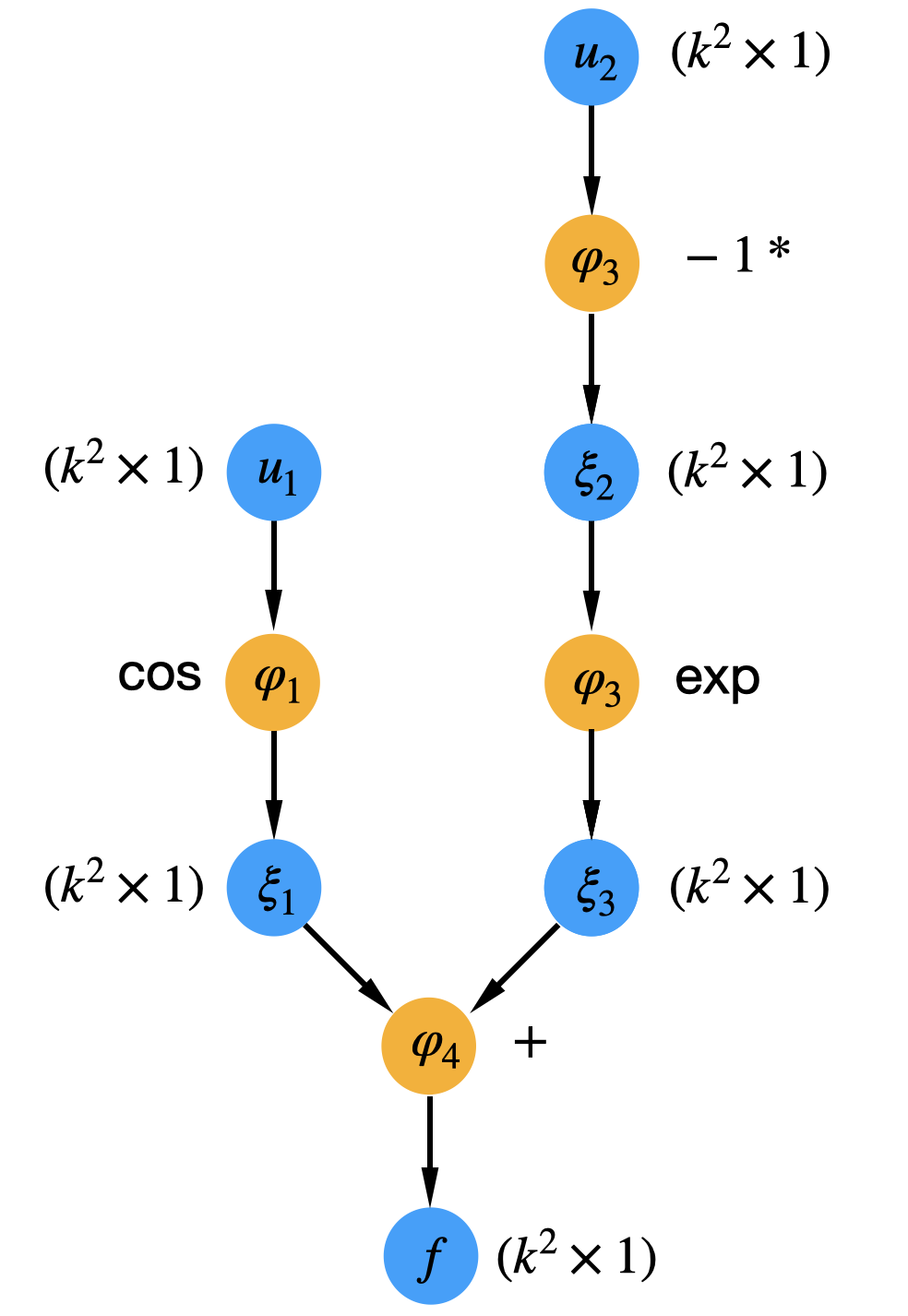} }}%
    \qquad
    \subfloat[\centering Computational graph using AMTC]{{\includegraphics[width=5cm]{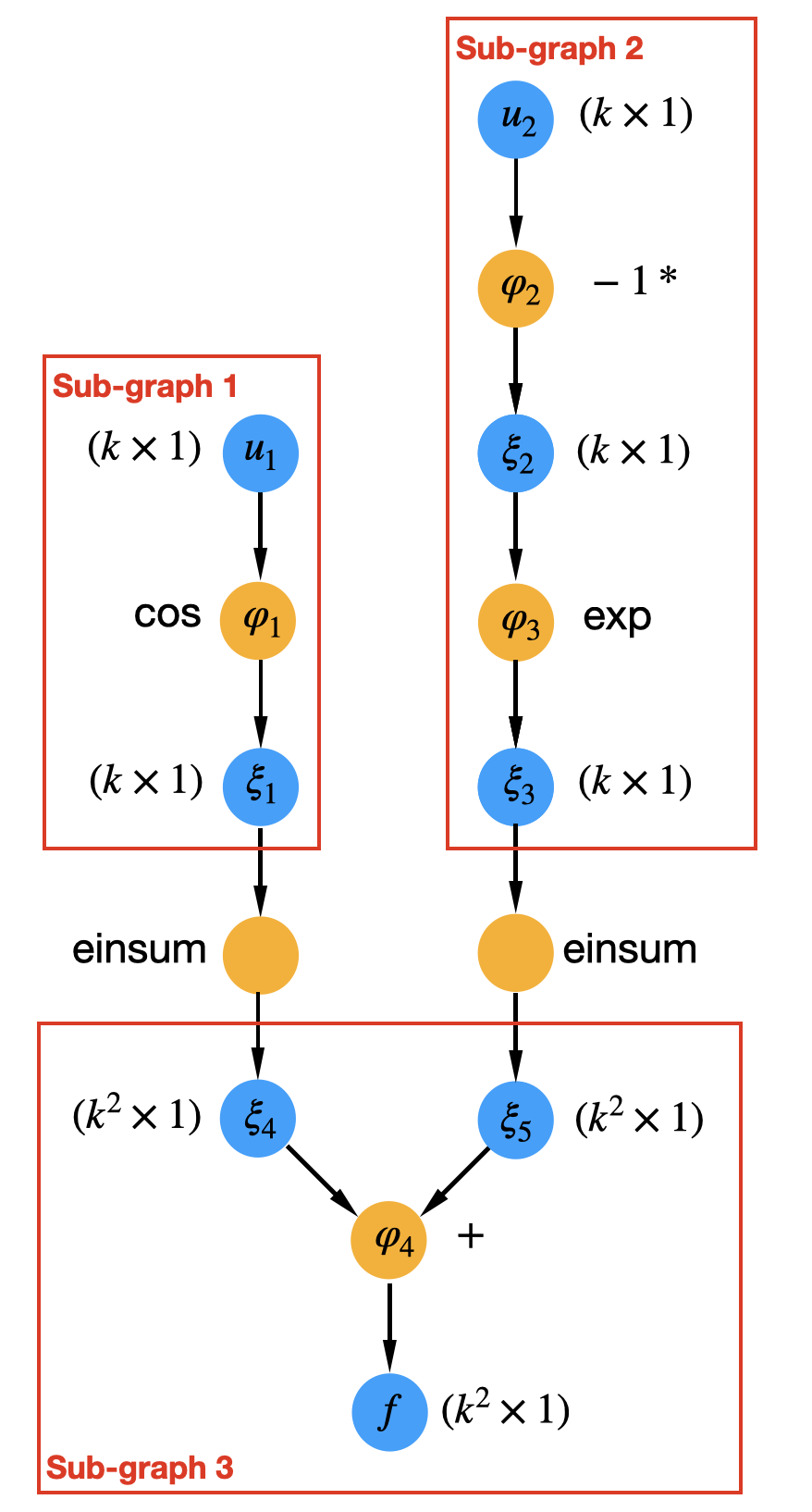} }}%
    \caption{Computational graphs with data size for full-grid input points evaluation on $f = cos(u_1) + exp(-u_2)$~\cite{wang2023accelerating}}%
    \label{fig:graph comparison}%
\end{figure}

Wang et al. recently proposed the \textit{Accelerated Model evaluations on Tensor grids using Computational graph transformations} (AMTC) method in
\cite{wang2023accelerating}.
AMTC is a graph transformation method to minimize the computational cost required to evaluate a computational model on tensor-grid input points. It represents the computational model as a computational graph built from basic operations and capitalizes on the fact that each operation only needs to be evaluated on the unique points in its input space. When we evaluate a computational model at tensor-grid input points, many operations depend on a small subset of the uncertain inputs. This leads to redundant evaluations at the operation level with a naive approach.

The AMTC method removes redundant calculations by partitioning the computational graph into smaller sub-graphs, based on the dependency information of the operations. Each sub-graph is computed just once based on the dependency information of the operations. 
Operations in each graph share the same input space and are evaluated only on the same distinct nodes in their input space. 
To maintain proper data flow among these sub-graphs, Einstein summation nodes are introduced to connect the sub-graphs. For illustration, take the function,
$f = \cos(u_1) + \exp(-u_2)$
evaluated at full-grid input points where each dimension has $k$ points. This function is broken down into fundamental operations:
\begin{equation}
\begin{aligned}
& \xi_1 = \varphi_1 (u_1) = \cos (u_1); \\
& \xi_2 = \varphi_2 (u_2) = - u_2; \\
& \xi_3 = \varphi_3 (\xi_2) = \exp (\xi_2); \\
& f = \varphi_4 (\xi_1, \xi_3) = \xi_1 + \xi_3. \\
\end{aligned}
\end{equation}
The tensor-grid input points are given as
\begin{equation}
\boldsymbol{u} = \begin{Bmatrix}
(u_1^{(1)}, u_2^{(1)}) & \ldots & (u_1^{(k)}, u_2^{(1)}) \\
\vdots & \ddots & \vdots \\
(u_1^{(1)}, u_2^{(k)}) & \ldots & (u_1^{(k)}, u_2^{(k)}) \\
\end{Bmatrix},
\end{equation}
with $k^2$ input points. 
The computational graphs to evaluate this function with and without the AMTC method are shown in Fig.~\ref{fig:graph comparison}.
Comparing the computational graphs reveals that without AMTC, every operation in the graph is evaluated $k^2$ times. But by applying AMTC, operations that depend solely on $u_1$ or $u_2$ are calculated just $k$ times, recognizing that their outputs have $k$ unique values. This method is especially beneficial when combined with the non-intrusive polynomial chaos method or quadrature rule for solving UQ problems when the model's computational graph is not densely connected.
AMTC has been integrated into the compiler's middle end for the \textit{Computational System Design Language} (CSDL)\cite{gandarillas2022novel}, a newly developed domain-specific language tailored for the solution of multidisciplinary design, analysis, and optimization problems. Within the CSDL compiler, the AMTC serves as a middle-end algorithm, generating the modified computational graph to minimize the model evaluation cost on tensor-grid inputs.

\section{Methodology}
\label{Sec: Methodology}

\subsection{Gradient-enhanced univariate dimension reduction}
We now describe the new methodology that forms the primary contribution of this paper.
The key idea here is to postulate a new approximation expression that enhances the accuracy of the univariate dimension reduction (UDR) approximation expression by adding the univariate gradient terms. 
With the addition of these gradient terms, the new approximation expression can be more accurate in estimating the second and higher-order statistical moments while the required computational cost still scales linearly with the problem dimension. 
We coin the new method as the \textit{gradient-enhanced univariate dimension reduction} (GUDR) method.
GUDR
approximates the original function, $f$, as
\begin{equation}
\begin{aligned}
    \bar{f}(u) & = \sum_{i=1}^d f_i(u_i) - (d-1) f(\mu) + \sum_{i=1}^d (u_i-\mu_i) \sum_{j \neq i}^d \frac{\partial f_j}{\partial{u_{i}}}(u_j)\\
    & - (d-1) \sum_{i}^d (u_i-\mu_i) \frac{\partial f}{\partial u_i}(\mu) - \sum_{i=1}^d \sum_{j >i}^d \frac{\partial^2 f }{\partial u_i \partial u_j} (\mu) (u_i-\mu_i)(u_j-\mu_j),
\end{aligned}
\end{equation}
where $\frac{\partial f_j}{\partial{u_{i}}}(u_j): = \frac{\partial f}{\partial{u_{i}}}(\mu_1, \ldots, \mu_{j-1}, u_j, \mu_{j+1}, \ldots, \mu_{d})$ represents a univariate gradient term.
This form includes all of the terms in the UDR approximation function in \eqref{eqn:udr}, with additional terms including the univariate first-order gradient terms and the second-order derivatives evaluated at $\mu$.
The GUDR approximation can also be written in matrix form as
\begin{equation}
\label{eqn: gudr}
\begin{aligned}
    \bar{f}(x) & = \sum_{i=1}^d f_i(u_i) - (d-1)f(\mu) +  \sum_{i=1}^d {\underbrace{\begin{bmatrix}
        u_1  - \mu_1 \\
        \vdots  \\
        u_{i-1} - \mu_{i-1} \\
        0  \\
        u_{i+1} - \mu_{i+1} \\
        \vdots \\
        u_d -\mu_d
    \end{bmatrix}}_{(u - \mu) \odot (\mathbf{1} -\mathbf{e_i})}}^T
    \underbrace{\begin{bmatrix}
        \frac{\partial f_i}{\partial u_1} (u_i)\\
        \vdots \\
        \frac{\partial f_i}{\partial u_d} (u_i)\\
    \end{bmatrix}}_{\frac{\partial f_i}{\partial u} (u_i) }\\
    & - (d-1) {\underbrace{\begin{bmatrix}
        u_1 -\mu_1 \\
        \vdots  \\
        u_d-\mu_d
    \end{bmatrix}}_{u-\mu}}^T     \underbrace{\begin{bmatrix}
        \frac{\partial f}{\partial u_1} (\mu)\\
        \vdots \\
        \frac{\partial f}{\partial u_d} (\mu)\\
    \end{bmatrix}}_{\frac{\partial f}{\partial u} (\mu)} -  {\underbrace{\begin{bmatrix}
     u_1 - \mu_1 \\
     \vdots \\
     u_d - \mu_d \\
    \end{bmatrix}^T \begin{bmatrix}
         0 & \frac{\partial^2 f}{\partial u_1 \partial u_2} (\mu) & \ldots &\frac{\partial^2 f}{\partial u_1 \partial u_d} (\mu) \\ 
         \vdots & \ddots & \ddots & \vdots \\
         0 & \ldots & 0& \frac{\partial^2 f}{\partial u_{d-1} \partial u_d} (\mu) \\
         0 & \ldots & \ldots  &  0
    \end{bmatrix}\begin{bmatrix}
     u_1 - \mu_1 \\
     \vdots \\
     u_d - \mu_d \\
    \end{bmatrix}}_{\frac{1}{2}(u-\mu)^T \left(\frac{\partial^2f}{\partial u^2}(\mu) - \text{diag}\left(\text{diag}
    \left(\frac{\partial^2f}{\partial u^2}(\mu)\right)\right)\right) (u-\mu)}} \\
   & = \sum_{i=1}^d f_i(u_i) +   \sum_{i=1}^d \left({(u - \mu) \odot (\mathbf{1} -\mathbf{e_i})}\right)^T \frac{\partial f_i}{\partial u} (u_i) - (d-1)\left(f(\mu) -(u-\mu)^T \frac{\partial f}{\partial u}(\mu)\right) \\
   &- \frac{1}{2}(u-\mu)^T \left(\frac{\partial^2f}{\partial u^2}(\mu) - \text{diag}\left(\text{diag}
    \left(\frac{\partial^2f}{\partial u^2}(\mu)\right)\right)\right) (u-\mu),
\end{aligned}
\end{equation}
where $\frac{\partial f}{\partial u}(\mu)$ and $\frac{\partial^2 f}{\partial u^2}(\mu)$ represent the gradient vector and Hessian matrix evaluated at $u = \mu$, respectively.
If we compare the Taylor series expansions at $u = \mu$ of the GUDR approximation function and the original function, just as with the UDR approximation, all of the Taylor series terms of the GUDR approximation function are included in the original function, and the residual errors can be expressed as
\begin{equation}
    f(u) - \bar{f}(u) = \sum_{i=1}^d\sum_{j > i}^d \sum_{k > j}^d \frac{\partial^3 f}{\partial u_i \partial u_j \partial u_k} (\mu)(u_i-\mu_i) (u_j - \mu_j) (u_k - \mu_k) +\ldots.
\end{equation}
This shows the GUDR approximation is third-order accurate when used to approximate the original function. In comparison with the residual errors of the UDR approximation in \eqref{eqn:general_taylor_UDR}, the GUDR approximation is a more accurate approximation of the original function as the UDR approximation is only second-order accurate.

When it comes to estimating the mean of the output, the GUDR approximation generates the same results as the UDR approximation with the residual errors expressed as
\begin{equation}
\begin{aligned}
\label{eqn:general_gudrerror_mean}
    \mathbb{E}[f(U)] -  \mathbb{E}[\bar{f}(U)] 
    & = \frac{1}{2!2!} \sum_{i_1 = 1}^{d-1} \sum_{i_2 > i_1}^{d}\frac{\partial^{4} f}{\partial u_{i_1}^{2} \partial u_{i_2}^{2}} (\mu) \mathbb{E}[(u_{i_1} -\mu_{i_1})^{2} (u_{i_2}-\mu_{i_2})^{2}] + \ldots.
\end{aligned}
\end{equation}
This means that the UDR and GUDR approximations are expected to provide the same results in estimating the mean of the output, and both are expected to be more accurate than the 3rd-order Taylor series expansion. However, when it comes to estimating the higher-order statistical moments, the GUDR can provide more accurate results than UDR as the GUDR approximation is more accurate than the UDR approximation in estimating the original function. 
For example, when estimating the second-order statistical moment of the output, the residual errors of the GUDR approximation can be expressed as

\begin{equation}
\label{eqn:general_gudrerror_sd}
     \mathbb{E}[f(U)^2] -  \mathbb{E}[\bar{f}(U)^2] =  \frac{1}{2!}f(\mu) \sum_{i_1 = 1}^{d-1} \sum_{i_2 > i_1}^{d}\frac{\partial^{4} f}{\partial u_{i_1}^{2} \partial u_{i_2}^{2}} (\mu) \mathbb{E}[(u_{i_1} -\mu_{i_1})^{2} (u_{i_2}-\mu_{i_2})^{2}] +\ldots
\end{equation}
When compared with the residual errors from the 3rd-order Taylor series expansion method, as detailed in \eqref{eqn:general_t3error_sd}, and those from the UDR method, seen in \eqref{eqn:general_udrerror_sd}, all three methods exhibit residual errors that include only terms of the fourth order and higher. 
However, the residual errors of the  GUDR and 3rd Taylor series expansion approximations contain the same fourth-order integration terms, while the UDR contains significantly more fourth-order integration terms.
This means that, when estimating the second-order statistical moments, the GUDR approximation is expected to have a comparable level of accuracy with the 3rd-order Taylor series expansion, and its result can be significantly more accurate than the UDR approximation.

The detailed derivation of the residual errors for the GUDR approximation in estimating the first and second-order statistical moments for a 2D problem can be found in Appendix~\ref{Sec: GUDR_2D}.

\subsection{Tensor-grid evaluations to estimate statistical moments}
For a $d$-dimensional UQ problem, the GUDR approximation in \eqref{eqn: gudr} involves multiple univariate function terms and univariate gradient terms. When it comes to estimating the mean of the output, all of the gradient terms become zero, and the GUDR yields the same result as UDR,
\begin{equation}
    \mathbb{E}[\bar{f}(U)] = \sum_{i=1}^d  \int_{\Gamma_i}
    f_i(u_i)\rho(u_i)du_i - (d-1) f(\mu),
\end{equation} 
which can be evaluated following the UDR method.
However, when it comes to estimating higher-order statistical moments of the output, there is no easy way to decompose the multidimensional integrals involved. 
One may want to derive the recursive formula for GUDR following the UDR method. However, the recursive formula will be significantly more complicated and is unpractical to implement in real problems.

In this paper, we propose a new approach to decompose the GUDR approximation function from the computational-graph-transformation perspective, so that we efficiently generate the model evaluations on full-grid quadrature points. 
This approach is inspired by the recently developed computational graph transformation method, \textit{Accelerated Model evaluations on Tensor-grid using Computational graph transformation} (AMTC).

\begin{figure}[hbt!]
\centering
  \includegraphics[width= 8cm]{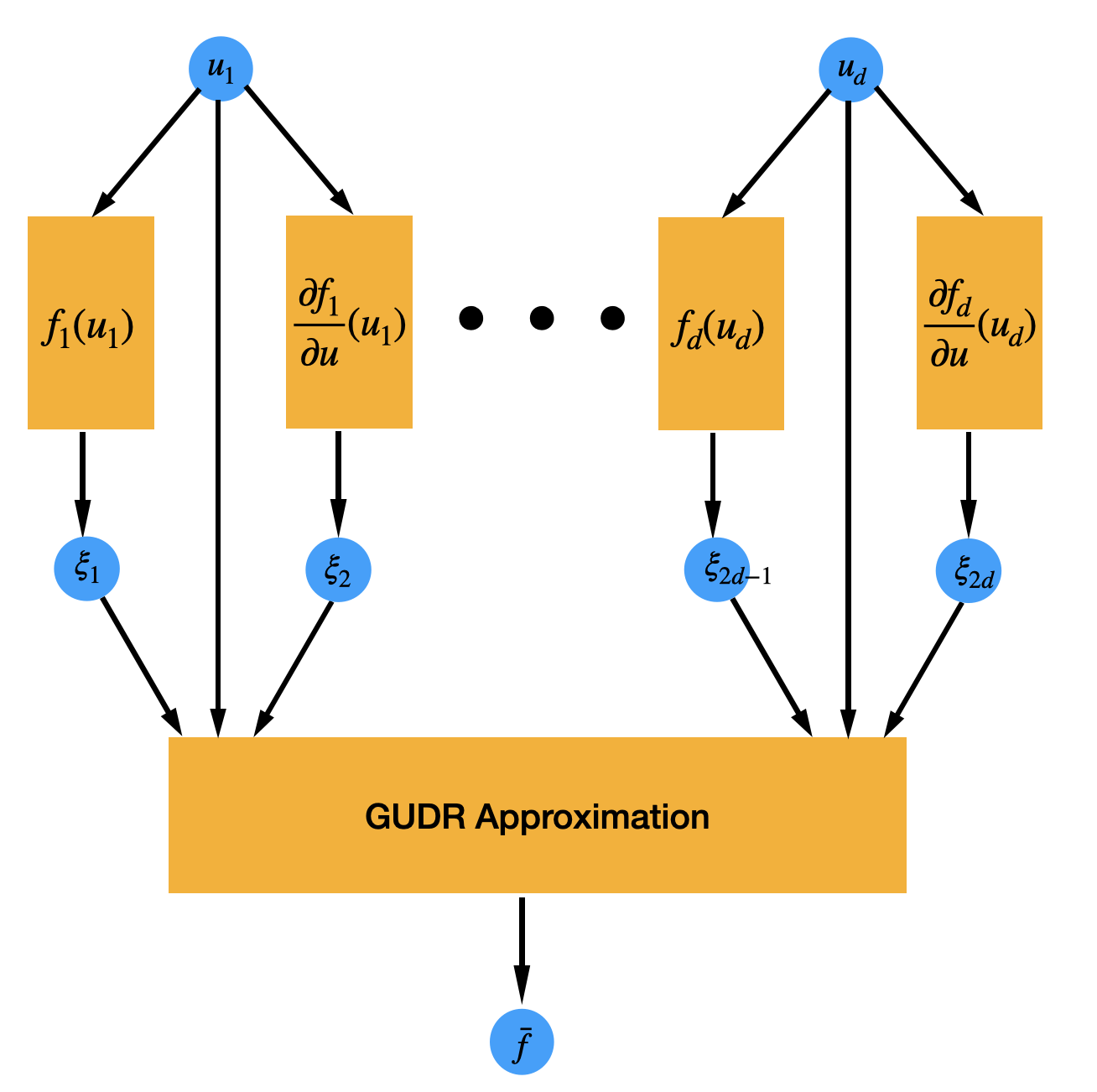}
\caption{Computational graph of the GUDR approximation function}
\label{fig:gudr_graph}
\end{figure}

If we treat each univariate function and gradient term in the GUDR approximation as individual operations, the computational graph of the GUDR approximation function is shown in Fig.~\ref{fig:gudr_graph}.
Despite the complicated form of the GUDR approximation, the main computational cost only comes from evaluating the univariate function and gradient terms.
When applying the GUDR approximation function to tensor-grid inputs, which are described as
\begin{equation}
\boldsymbol{u} =
\boldsymbol{u}^{k}_{{1}}
\times \ldots \times
\boldsymbol{u}^{k}_{{d}},
\end{equation}
where $\boldsymbol{u}^{k}_{{i}}$ denotes the set of $k$ quadrature points within the $u_{i}$ dimension, we note that despite the presence of $k^d$ input points, only $k$ distinct points exist within each input dimension. 
Drawing on AMTC's foundational concept, since every univariate function and gradient computation operation in the GUDR's computational graph is dependent on just one uncertain input, it is only necessary to evaluate these operations at the distinct quadrature points within their input space.

\begin{figure}[hbt!]
\centering
  \includegraphics[width= 10cm]{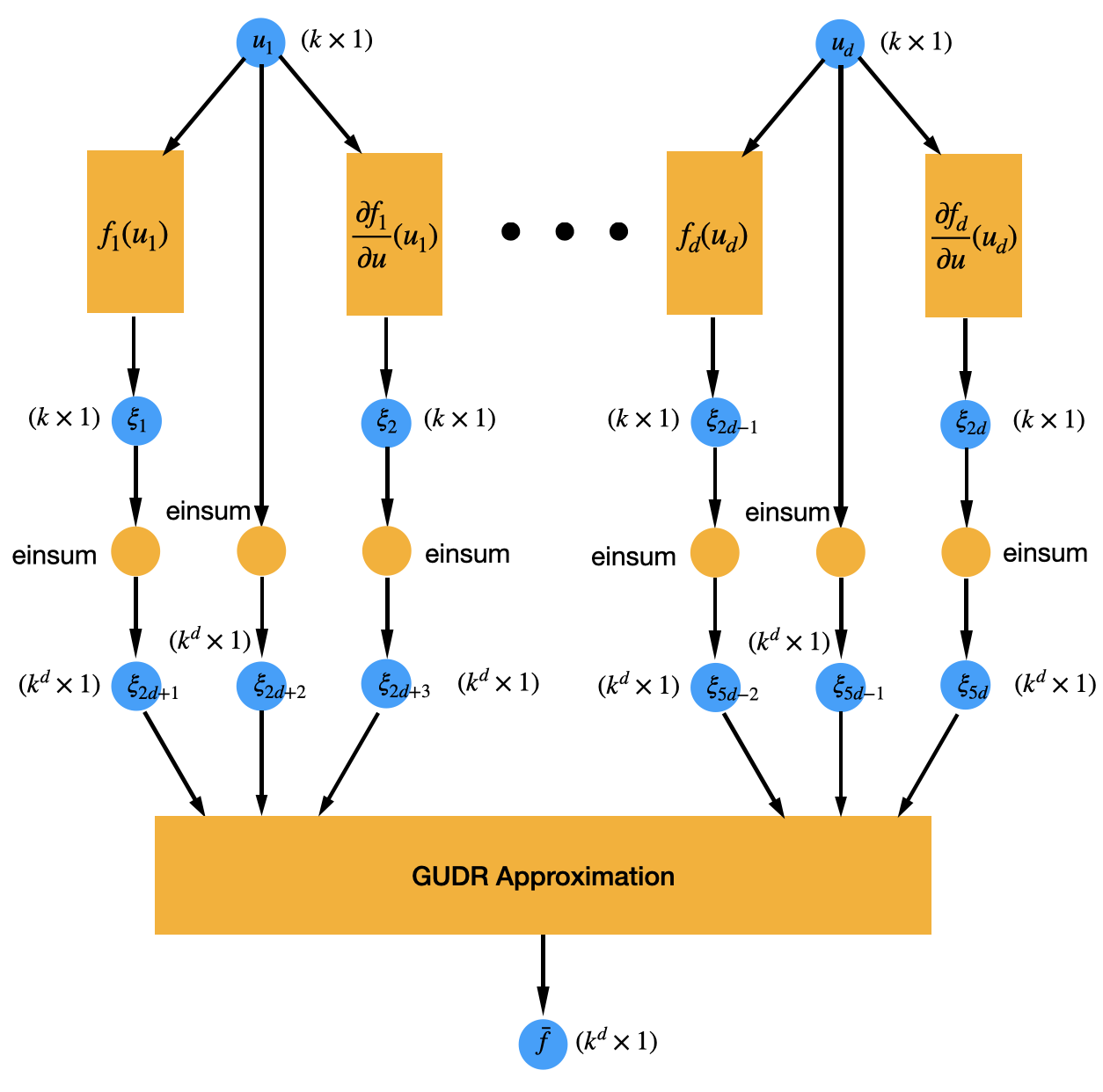}
\caption{Computational graph of tensor-grid evaluations after graph transformation}
\label{fig:gudr_amtc}
\end{figure}

Consequently, by integrating the AMTC strategy, the modified computational graph for the GUDR approximation function's tensor-grid evaluations is shown in Fig. \ref{fig:gudr_amtc}.
In this way, we can generate the model evaluations of GUDR approximation function on tensor-grid inputs, $\bar{f}(\boldsymbol{u}) \in \mathcal{R}^{k^d}$, with only $k$ evaluations on each univariate function and gradient evaluation term.

We note that the AMTC method has only been implemented in the CSDL compiler and can only be applied to computational models that are built in the CSDL language.
Fortunately, in this specific scenario, we can achieve the minimum number of evaluations in univariate function and gradient evaluation terms by manually adding the \textit{Einsum}\footnote{Einsum refers to the Einstein summation function in Numpy.} operations.
First, we evaluate the univariate model function and gradient function on the corresponding quadrature points in its input space. Then we manually add \textit{Einsum} operations in the code to transform these quadrature points evaluations to the correct size with the correct order of the data. Lastly, this data is passed to the GUDR approximation function to generate the full-grid evaluations of the output. We show example codes to achieve this in the Appendix~\ref{sec: python_code}.

The tensor-grid input-output data can be easily used with the quadrature rule to compute any order of statistical moments of the output.
Additionally, this data can also be used with other UQ methods like non-intrusive polynomial chaos, stochastic collocation, and kriging to construct a surrogate model to approximate the risk measures of the output, such as the probability of failure or conditional value at risk.

\subsection{Computational cost}
Utilizing the AMTC strategy enables the computation of any statistical moment or risk measure of the output with a fixed number of evaluations. Specifically, GUDR would only require $k$ evaluations for each univariate function and its gradient term, accompanied by a single function evaluation, gradient evaluation, and Hessian evaluation at the mean of the input variables. 
With reverse-mode automatic differentiation, for a function with $d$ inputs,  a single gradient vector evaluation, and a full Hessian matrix evaluation carry the computational weight less than 3 and  $3d$ model evaluations~\cite{christianson1992automatic}, respectively.
As a result, a conservative estimation of the model evaluation cost for solving the $d$-dimensional UQ using GUDR with $k$ quadrature points in each dimension of the uncertain input is no more than $4kd + 3d + 4$ model evaluations, which scales only linearly with the dimension of the problem.

\section{Numerical Results}
\label{Sec: Numerical Results}
In Sec \ref{Sec:math_func}, we compare the accuracy of four UQ methods for estimating the standard deviation of the output: gradient-enhanced univariate dimension reduction (GUDR), univariate dimension reduction (UDR), method of moments using a 2nd order Taylor series expansion, and method of moments using a 3rd order Taylor series expansion.
These methods are applied to two UQ problems involving different mathematical functions.
In Sec \ref{Sec:math_func}, we compare the scalability of three UQ methods: UDR, GUDR, and full-grid quadrature rule. These methods are applied to one UQ problem involving a mathematical function with varying dimensions. 
In Sec \ref{Sec:bem} and Sec \ref{Sec:aircraft}, we compare the convergence plots of GUDR, UDR, and the Monte Carlo methods in estimating both the mean and the standard deviation of the output on a 4D UQ problem involving a rotor aerodynamic analysis model and a 7D UQ problem involving an aircraft design multidisciplinary model, respectively.

\subsection{Mathematical functions {with varying input variances}}
\label{Sec:math_func}
We first examine two UQ problems that involve mathematical functions defined by simple, closed-form expressions.
The first problem involves the function,
\begin{equation}
y_1 = f(x_1,x_2) = \frac{1}{1 + x_1^4 + 2 x_2^2 + x_2^4}
\end{equation}
with uncertain inputs,
\begin{equation}
X_i \sim \mathcal{N}(2,\sigma), \quad i = 1,2.
\end{equation}
The second problem involves the function:
\begin{equation}
y_2 = f(x_1,x_2,x_3) = \exp(1+0.5x_1^2 + 0.5x_2^2+ 0.5x_3^2)
\end{equation}
with uncertain inputs:
\begin{equation}
X_i \sim \mathcal{N}(3,\sigma), \quad i = 1,2,3.
\end{equation}

\noindent
The quantity of interest (QoI) for both UQ problems is the standard deviation of the output $Y_1$ and $Y_2$, respectively. These problems are adapted from \cite{rahman2004univariate}.

For these test problems, we have implemented five methods: UDR, GUDR, method of moments using the 2nd-order Taylor series expansion (second-order second-moment (SOSM)), method of moments using the 3rd-order Taylor series expansion (third-order second-moment (TOSM)), and direct numerical integration. The UDR and GUDR methods are implemented using 19 quadrature points in each dimension of the uncertain inputs to yield the most accurate UQ results based on these two analytical approximation expressions. Conversely, the direct numerical integration method utilizes the Monte Carlo method with 100,000 sample points, with its UQ results considered as the ground truth.

\begin{figure}
    \centering
    \subfloat[\centering QoI results for increasing values of input standard deviation]{{\includegraphics[width=7cm]{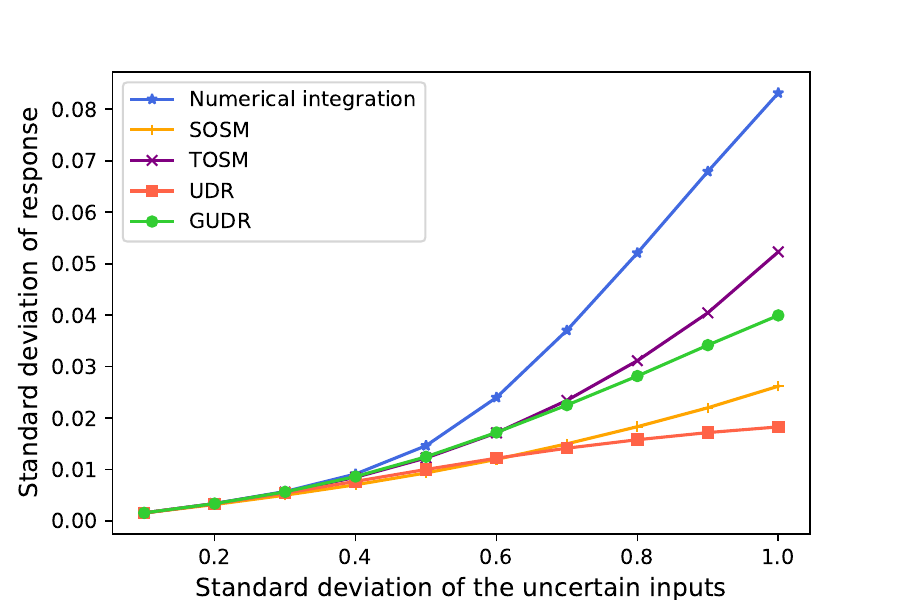} }}
    \qquad
    \subfloat[\centering Relative errors for increasing values of input standard deviation]{{\includegraphics[width=7cm]{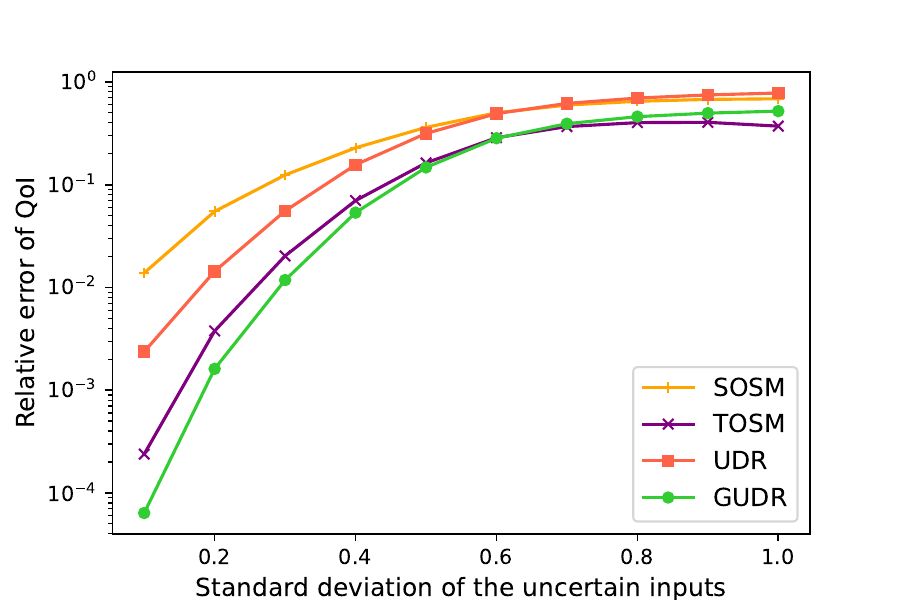}}}
    \caption{UQ results for $ y_1 = (1 + x_1^4 + 2 x_2^2 + 5_2^4)^{-1}$ }%
    \label{fig:f1_result}
\end{figure}

\begin{figure}%
    \centering
    \subfloat[\centering QoI results for increasing values of input standard deviation]{{\includegraphics[width=7cm]{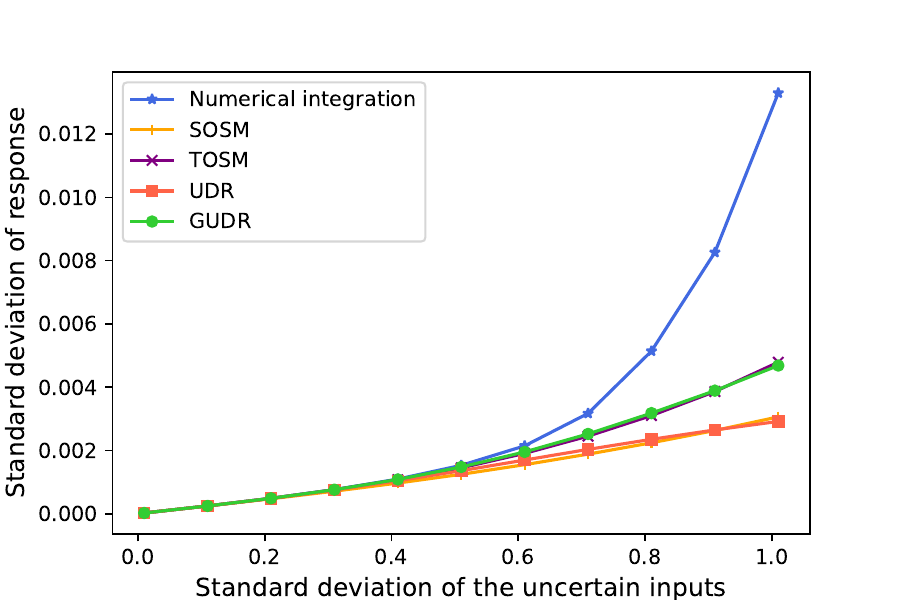} }}
    \qquad
    \subfloat[\centering Relative errors for increasing values of input standard deviation]{{\includegraphics[width=7cm]{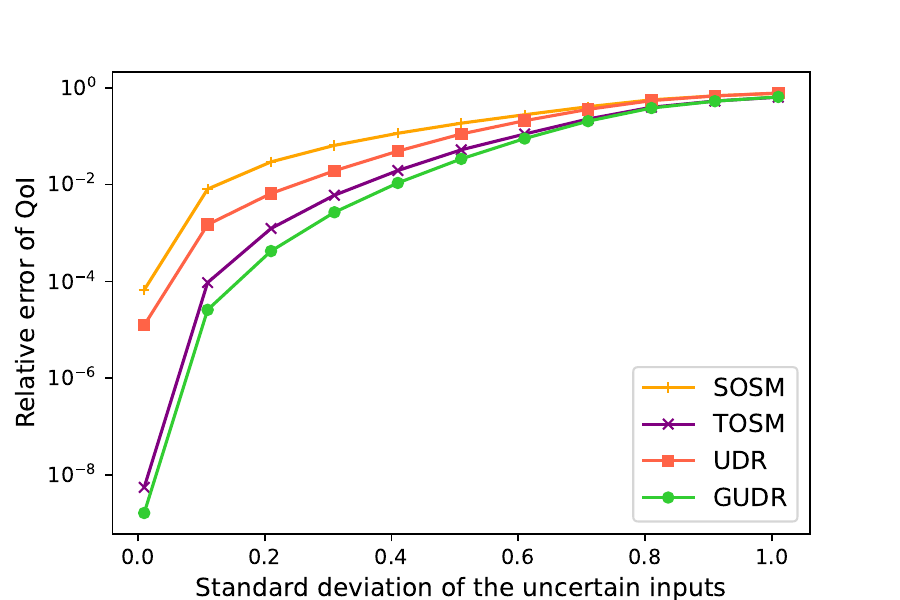}}}
    \caption{UQ results for $  y_2 = e^{(1+0.5x_1^2 + 0.5x_2^2+ 0.5x_3^2)}$ }%
    \label{fig:f2_result}%
\end{figure}

The UQ results derived from these five methods, along with the relative errors of the four analytical expression methods, are plotted against the increasing values of input standard deviations in Fig.~\ref{fig:f1_result} and Fig.~\ref{fig:f2_result} for $Y_1$ and $Y_2$, respectively. 
Our observations indicate 
that for both problems, all of the UDR, GUDR, SOSM, and TOSM methods yield accurate results for the standard deviation of the response when the standard deviation of the uncertain inputs is small. As we increase the standard deviation, the errors for all four methods correspondingly escalate. 
While the performances of UDR and SOSM are comparable, GUDR and TOSM also have comparable performances and consistently outperform UDR and SOSM. Typically, the relative errors of GUDR are less than those of UDR and SOSM by more than an order of magnitude. This superior performance of GUDR can be attributed to its more accurate approximation of the original function in comparison to UDR and the 2nd-order Taylor expansion. Additionally, the comparable performances of GUDR and TOSM match our theoretical results showing that the GUDR approximation function is expected to have comparable performance as the 3rd-order Taylor expansion when estimating the standard deviation of the output.

\subsection{{Mathematical function with varying dimensions}}
{The third test problem involves an analytical problem with varying dimension:
\begin{equation}
y_3 = f(x_1,\ldots, x_d) = 
\exp(1+ \sum_{i=1}^d 0.5x_i^2)
\end{equation}
with uncertain inputs:
\begin{equation}
X_i \sim \mathcal{N}(3,1), \quad i = 1,\ldots, d.
\end{equation}
For this test problem, we aim to investigate how the function evaluation cost scales with the dimension of the function for the GUDR method. We have implemented three methods: UDR, GUDR and direct numerical integration with full-grid quadrature rule. For all three methods, we use $k = 5$ as the number of quadrature points in each dimension. For UDR and GUDR, the derivatives are computed using the automatic differentiation from TensorFlow.
The function evaluations time for each case is measured as the average of 200 runs.}

\begin{figure}[hbt!]
\centering
  \includegraphics[width= 8cm]{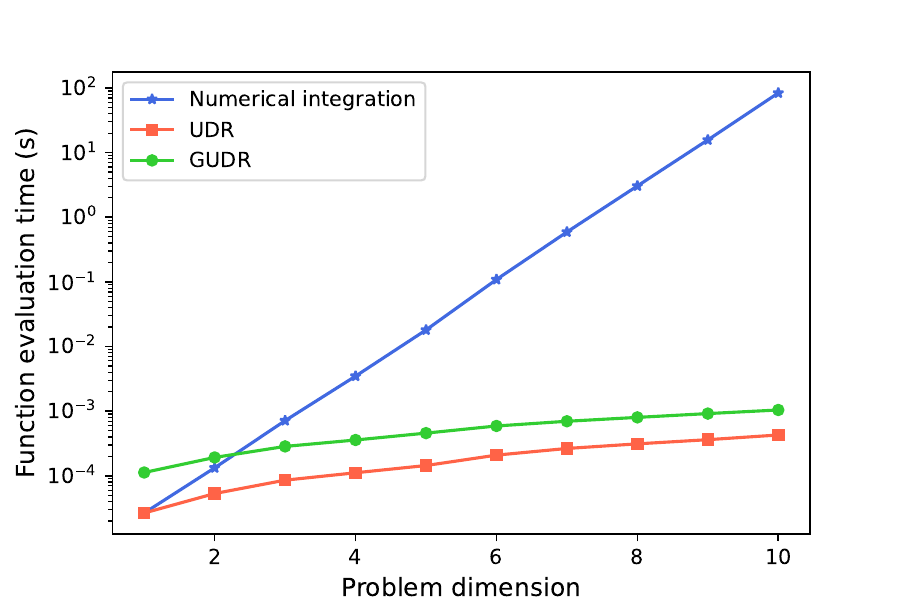}
\caption{{Scalability plots for $y_3 = \exp(1+ \sum_{i=1}^d 0.5x_i^2)$}}
\label{fig:scalability_plot}
\end{figure}

{The scalability results are plotted inFig.~\ref{fig:scalability_plot}. Our observations indicate that both UDR and GUDR show linear scalability with the problem dimension, while numerical integration with the full-grid quadrature rule exhibits exponential scalability. Comparing GUDR with UDR, the results show that the computational cost of GUDR is consistently larger than that of UDR, with a ratio between 2 and 4. These observations align with our theoretical analysis. When using an efficient automatic differentiation method to compute the gradients, both GUDR and UDR scale linearly with the problem dimension, and the computational cost of GUDR is estimated to be no more than four times that of UDR.}

\subsection{Rotor aerodynamic analysis model}
\label{Sec:bem}

\begin{figure}[hbt!]
\centering
  \includegraphics[width= 10cm]{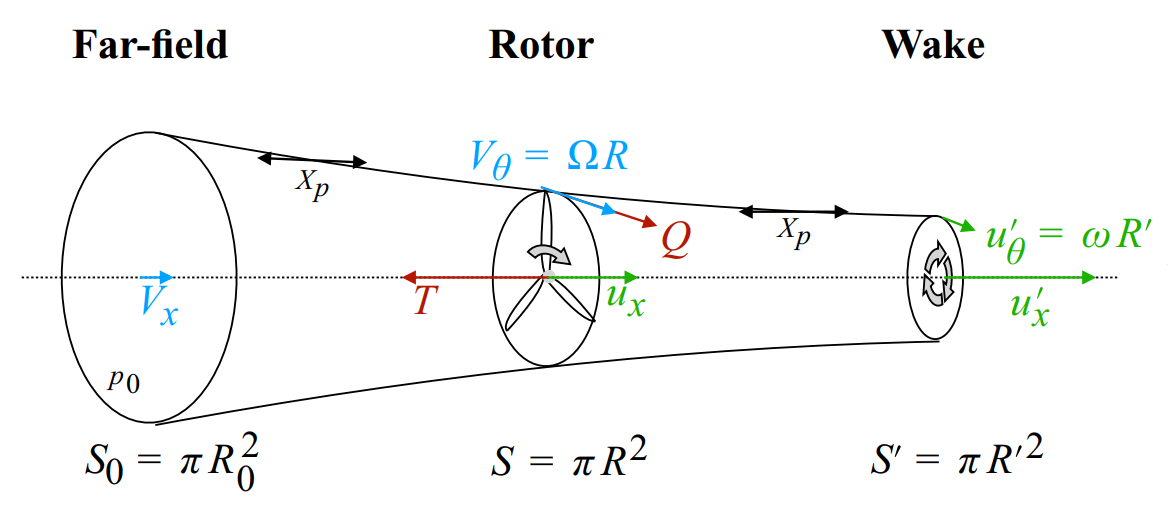}
\caption{Streamtube for the blade element momentum model~\cite{ruh2023fast}}
\label{fig:bem_demo}
\end{figure}

The {fourth} test problem is a 4D UQ problem involving the aerodynamic analysis of a rotor on an aircraft. The computational model applies the blade element momentum (BEM) theory to compute the rotor performance metrics including the thrust and torque generated. The streamtube analyzed by the BEM model is shown in Fig.~\ref{fig:bem_demo}. Further details of the model can be found in~\cite{ruh2023fast}. The objective of the UQ problem is to determine the mean and standard deviation of the generated torque, given four uniform uncertain inputs shown in Tab.~\ref{tab:beam_uncertain_inputs}.

\begin{figure}%
    \centering
    \subfloat[\centering Mean of the output]{{\includegraphics[width=7cm]{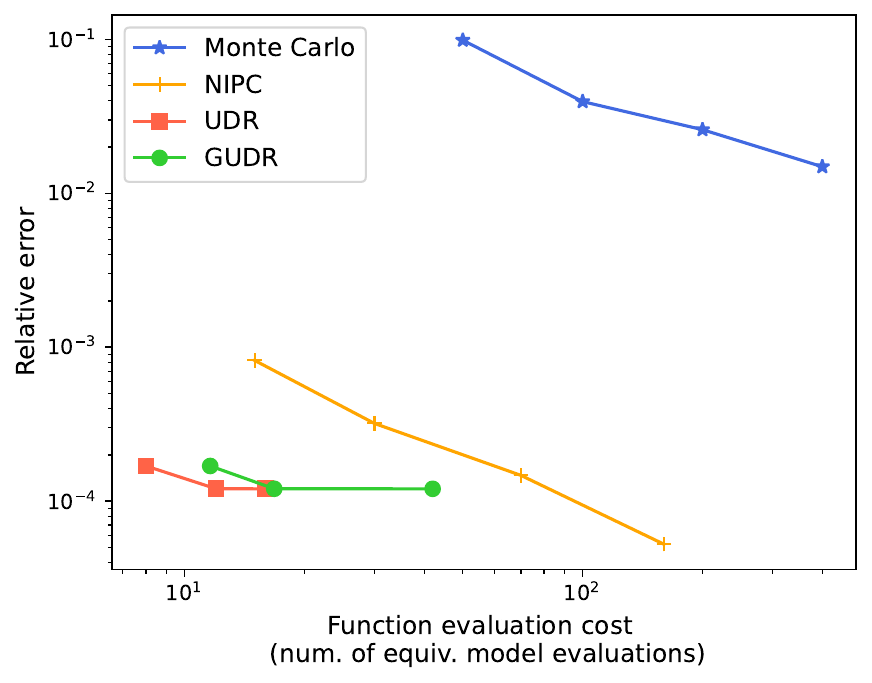} }}
    \qquad
    \subfloat[\centering Standard deviation of the output]{{\includegraphics[width=7cm]{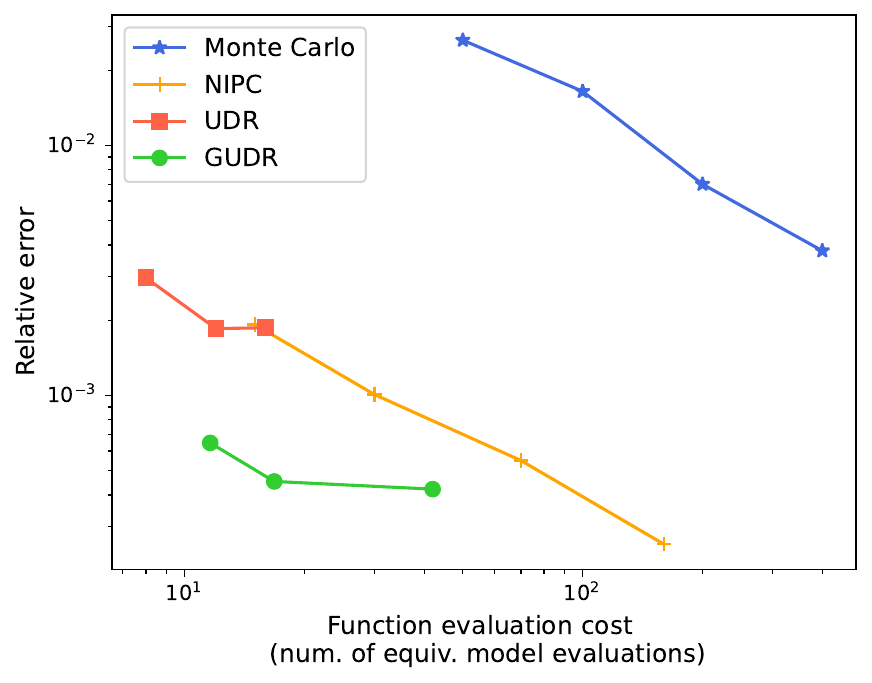}}}
    \caption{Convergence plots for the rotor analysis problem}%
    \label{fig:bem_result}%
\end{figure}

In this and the following test problems, we implemented four UQ methods: GUDR, UDR, non-intrusive polynomial chaos (NIPC), and the Monte Carlo method. 
The computational models were implemented in the Computational System Design Language (CSDL) \cite{gandarillas2022novel}. The first-order gradient evaluations required for GUDR were calculated using an automatically implemented adjoint method detailed in \cite{sperry2023automatic}, and the Hessian matrix was estimated via a finite difference method. The NIPC results were generated using the software package Chaospy~\cite{feinberg2015chaospy}, using a regression-based approach.
We use the UQ results generated from the Monte Carlo method with $10,000$ sample points as the reference results and depict the convergence plots of the four methods in Fig.~\ref{fig:bem_result}.
{For this and the following test problems, the function evaluation cost for each data point is computed as the total evaluation time (including both function and gradient evaluations) divided by the computational time of a single function evaluation, represented as function evaluation cost (num. of function evaluations).}

\begin{table}[h!]
    \centering
    \begin{tabular}{c | c }
         Uncertain inputs & Distributions \\
         \hline
         \text {Rotor speed (RPM)} & {$U(1600, 1800)$} \\
         \text {Axial free-stream velocity (m/s)} & {$U(50, 60)$} \\
         \text {Propeller radius (m)} & {$U(0.79, 0.81)$} \\
         \text {Flight altitude (m)} & {$U(9000, 10000)$} \\
    \end{tabular}
    \caption{Input parameters and ranges for the rotor analysis problem}
    \label{tab:beam_uncertain_inputs}
\end{table}

From the results, it is evident that at a low evaluation cost (10s of equivalent model evaluations), UDR, GUDR, and NIPC are capable of delivering more accurate results compared to the Monte Carlo method in terms of both mean and standard deviation for this particular problem. 
In terms of estimating the mean of the output, UDR and GUDR yield identical UQ results when employing the same number of quadrature points in each dimension, but GUDR is computationally more demanding than UDR. 
This observation is aligned with our theoretical analysis, which affirms that both GUDR and UDR are fourth-order accurate and produce identical results when estimating the mean of the output. 
Additionally, both UDR and GUDR are more accurate than NIPC at low evaluation costs (10s of equivalent model evaluations).

As for the standard deviation of the output, UDR's accuracy is comparable to NIPC, while GUDR enhances the accuracy of UDR by almost an order of magnitude at the expense of a slightly increased computational cost. 
This phenomenon can be attributed to the fact that the GUDR approximation function results in fewer relative errors than the UDR approximation function when estimating the second and higher-order statistical moments. 
In this problem, with 20 equivalent model evaluations, GUDR can achieve a relative error of less than $0.1\%$ for both mean and standard deviation of the output, making the GUDR method the most affordable UQ method to achieve this level of performance.

\subsection{Aircraft design multidisciplinary model}
\label{Sec:aircraft}

\begin{figure}[hbt!]
\centering
  \includegraphics[width= 10cm]{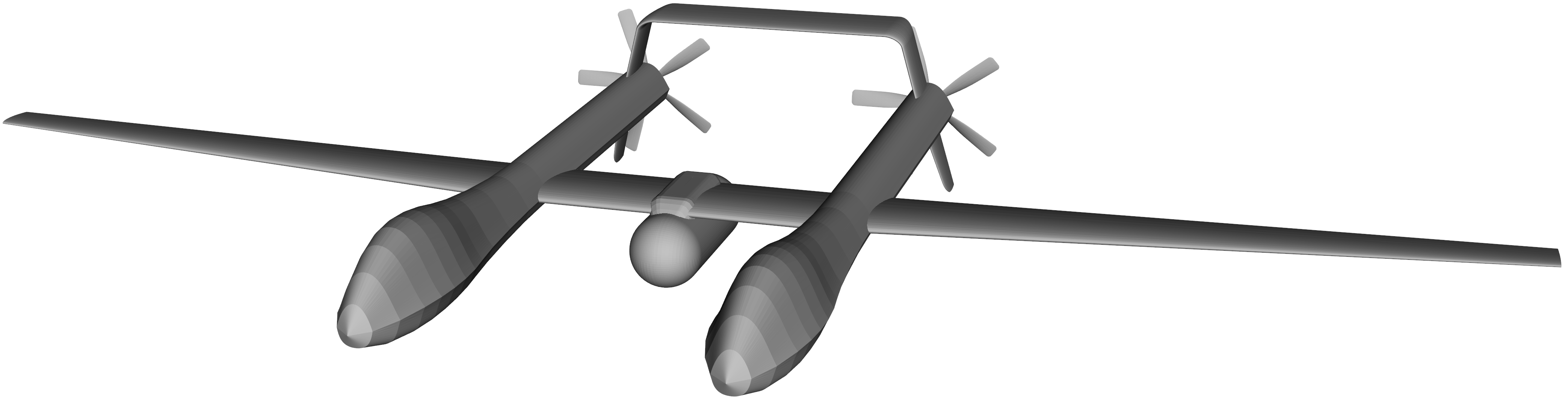}
\caption{The high-altitude, laser-powered airplane concept}
\label{fig:uav_concept}
\end{figure}

\begin{figure}[hbt!]
\centering
  \includegraphics[width= 6cm]{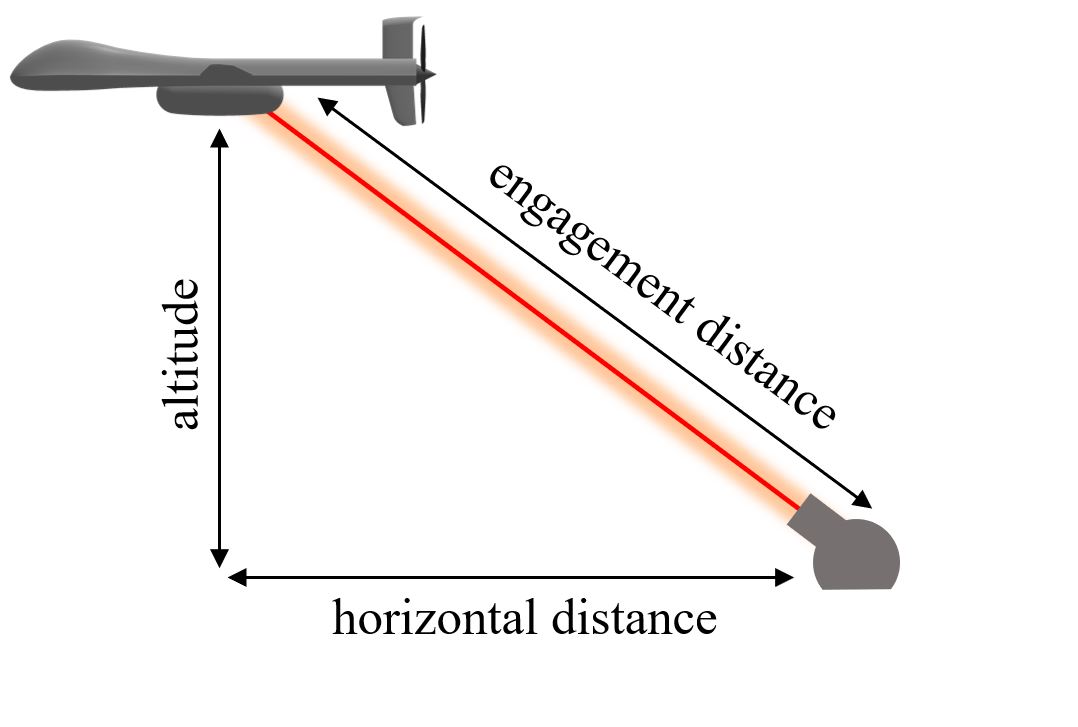}
\caption{The laser beam engagement scenario}
\label{fig:laser}
\end{figure}

\begin{figure}
    \centering
    \subfloat[\centering 1D beam meshes for the structure solver]{{\includegraphics[width=6cm]{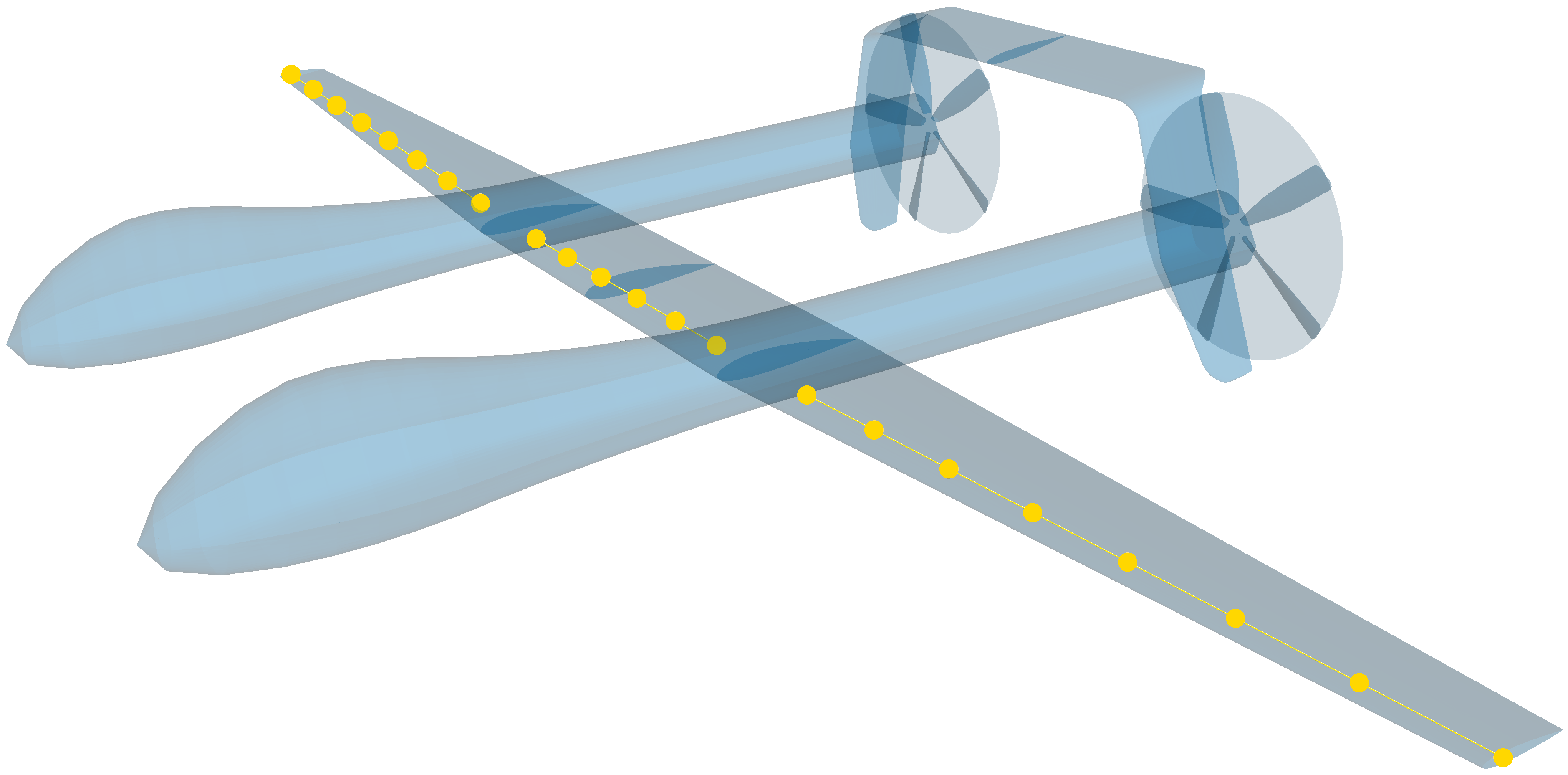} }}
    \qquad
    \subfloat[\centering 2D aerodynamics meshes for the aerodynamics solver]{{\includegraphics[width=6cm]{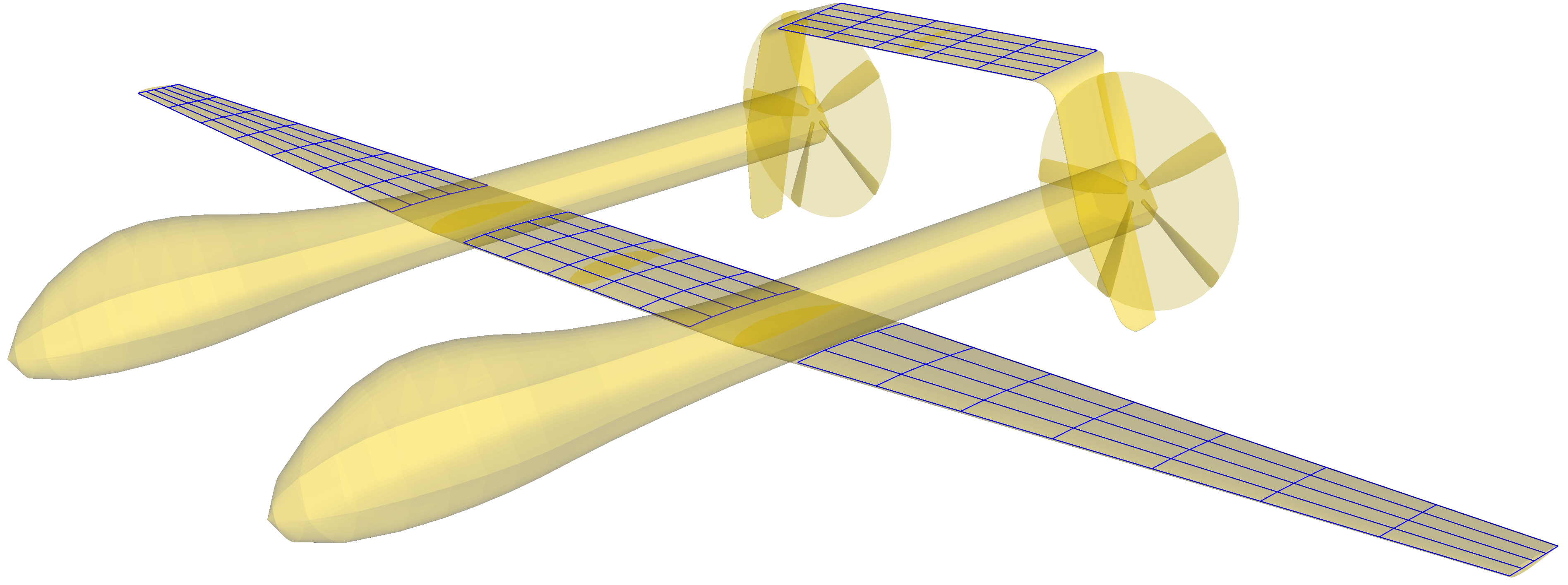}}}
    \caption{Meshes for structure and aerodynamics solvers}
    \label{fig:meshes}
\end{figure}

{fifth} test problem is a 7D UQ problem derived from a practical aircraft design scenario. This problem involves a high-altitude, laser beam-powered unmanned aerial vehicle (UAV) cruising at a high altitude while receiving charge from a laser beam emitted by a ground station. 
The concept of the UAV is illustrated in Fig.~\ref{fig:uav_concept}, while the laser beam engagement scenario is depicted in Fig.~\ref{fig:laser}. 
The computational model incorporates multiple disciplines, such as aerodynamics, structures, thermals, power beaming, performance, and weight models. This model calculates the required aircraft mass to meet certain criteria, including endurance, maximum stress, and static margin. The meshes for the aerodynamics and structure solvers are shown in Fig.~\ref{fig:meshes}. For further details on the computational model, we refer the reader to \cite{orndorff2023gradient}.

\begin{figure}%
    \centering
    \subfloat[\centering Mean of the output]{{\includegraphics[width=7cm]{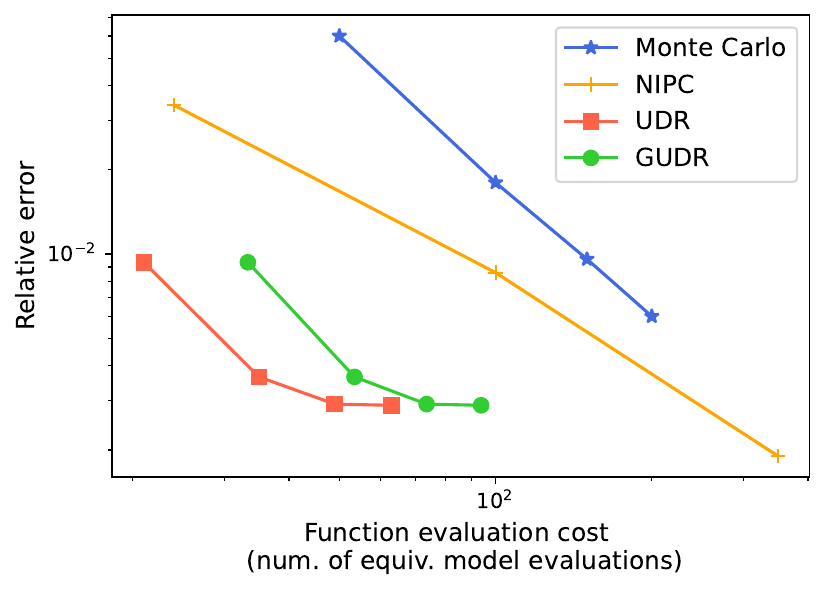} }}
    \qquad
    \subfloat[\centering Standard deviation of the output]{{\includegraphics[width=7cm]{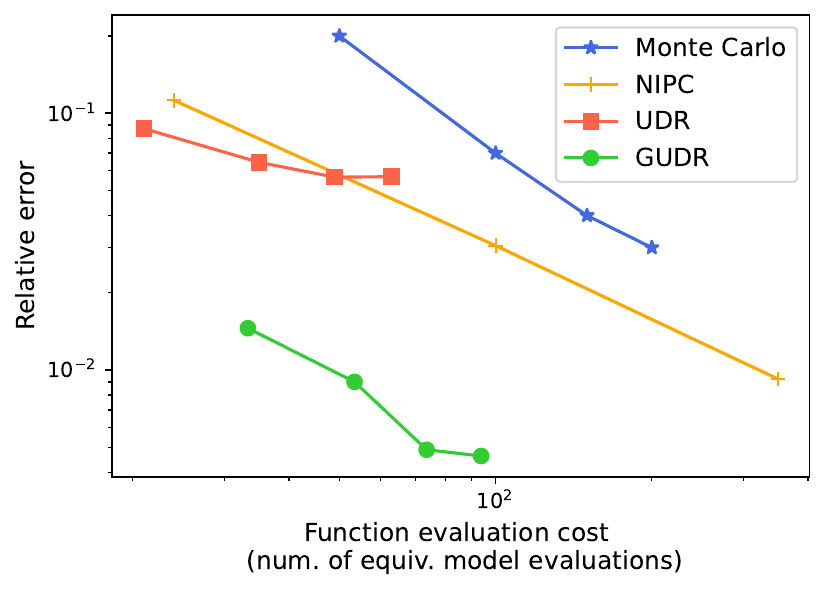}}}
    \caption{Convergence plots for the aircraft design problem}%
    \label{fig:darpa_result}%
\end{figure}

The objective of the UQ problem is to determine the mean and standard deviation of the required aircraft mass, taking into account seven uncertain inputs with uniform distributions. The specific uncertain inputs and their distributions are listed in Tab.~\ref{tab:uncertain_inputs}.
{We use the UQ results generated from the Monte Carlo method with $10,000$ sample points as the reference results and depict the convergence plots of the four methods in Fig.~\ref{fig:darpa_result}.}


\begin{table}[h!]
    \centering
    \begin{tabular}{c | c }
         Uncertain inputs & Distributions \\
         \hline
         \text {Cruise mach number} & {$U(0.1, 0.2)$} \\
         \text {Altitude}(km) & {$U(17, 19)$} \\
         \text {Horizontal distance} (km) & {$U(250, 300)$} \\
         \text {Pitch angle} (deg) & {$U(-2, 2)$} \\
         \text {Payload mass} (kg) & {$U(50, 100)$} \\
         \text {Aperture diameter} (m) & {$U(0.7, 0.8)$} \\
         \text {Rotor speed } (RPM) & {$U(1200, 1500)$}  \\
    \end{tabular}
    \caption{Input parameters and ranges for the aircraft design problem}
    \label{tab:uncertain_inputs}
\end{table}

Similar to the last test problem, the Monte Carlo method is outperformed by UDR, GUDR, and NIPC at low evaluation costs.
When estimating the mean of the output, GUDR is slightly more expensive than UDR and both of them are more accurate than NIPC.
However, when estimating the standard deviation of the output, UDR has comparable performance to NIPC at low function evaluation costs. In contrast, GUDR enhances the result of UDR by almost an order of magnitude with a slight increase in computational cost. 
These observations are also consistent with our theoretical results as GUDR can be significantly more accurate than UDR in estimating second and higher-order statistical moments of the output.
As a result, in this problem, with the computational cost of 50 equivalent model evaluations, GUDR can achieve a relative error of less than $1\%$ for both the mean and standard deviation of the output, making the GUDR method the most affordable UQ method for achieving this level of performance.

\section{Conclusion}
\label{Sec: Conclusion}
In this paper, we introduced a UQ method, \textit{gradient-enhanced univariate dimension reduction} (GUDR), for solving high-dimensional UQ problems.
This method builds on the idea of univariate dimension reduction (UDR) and adds univariate gradient functions into the UDR approximation function so that it enhances its accuracy in approximating the original function.
GUDR employs a computational graph transformation strategy to efficiently evaluate its approximation function on tensor-grid quadrature points to evaluate any statistical moment of the output.
By using an automatic differentiation method to evaluate the gradient information, GUDR preserves the linear scaling cost of the UDR method.

The GUDR method was tested on {five} UQ problems: two mathematical functions in 2D and 3D, one mathematical function with varying dimensions, one 4D rotor analysis problem, and one 7D  aircraft design problem.
For the {first two} mathematical functions, 
GUDR surpassed UDR in accuracy for standard deviation estimation and matched the performance of the method of moments using the 3rd-order Taylor expansion.
{For the third mathematical function, GUDR exhibits linear scalability with the problem dimension in terms of function evaluation cost, which is less than four times that of UDR.}
In the 4D rotor analysis and 7D aircraft design problems, GUDR improved upon UDR's precision tenfold when estimating the output's standard deviation, achieving the most cost-effective performance among the UQ methods implemented.

A critical limitation of this work is that GUDR requires the evaluations of both first- and second-order gradients of the computational model using an automatic differentiation method, which is not a universal feature in all software environments. 
{One important direction for future work is to evaluate the GUDR's performance in estimating third and higher-order statistical moments, such as skewness and kurtosis. Another promising area is to assess the GUDR's accuracy in estimating reliability metrics of the output by integrating it with polynomial chaos and kriging methods. Additionally, developing more accurate linear-scaling UQ methods based on the GUDR concept presents an interesting research opportunity. Furthermore, another important research question is how to develop a method that improves the accuracy of UDR in estimating the mean of the output by incorporating gradient information.}

\section*{Appendix}
\subsection{Univariate dimension reduction in 2D case}
\label{Sec: UDR_2D}
In this section, we derive the residual errors of the univariate dimension reduction (UDR) method in estimating the mean and standard deviation for a 2D UQ problem.  
Consider an UQ problem that involves a bivariate function $f\left(u_1, u_2\right)$, the Taylor series expansion of $f$ at ($u_1 = \mu_1, u_2 = \mu_2$) can be expressed by
\begin{equation}
\label{eqn:2d_taylor_original}
    \begin{aligned}
f\left(u_1, u_2\right)= & f(\mu_1,\mu_2)+\frac{\partial f}{\partial u_1}(\mu_1,\mu_2) (u_1-\mu_1)+\frac{\partial f}{\partial u_2}(\mu_1,\mu_2) (u_2-\mu_2) 
 +\frac{1}{2 !} \frac{\partial^2 f}{\partial u_1^2}(\mu_1,\mu_2) (u_1-\mu_1)^2 \\
 &+\frac{1}{2 !}\frac{\partial^2 f}{\partial u_2^2}(\mu_1,\mu_2) (u_2-\mu_2)^2  + \frac{\partial^2 f}{\partial u_1 \partial u_2}(\mu_1,\mu_2) (u_1-\mu_1) (u_2-\mu_2) 
 +\frac{1}{3 !} \frac{\partial^3 f}{\partial u_1^3}(\mu_1,\mu_2) (u_1-\mu_1)^3 \\
 & +\frac{1}{3 !} \frac{\partial^3 f}{\partial u_2^3}(\mu_1,\mu_2) (u_2-\mu_2)^3+\frac{1}{2 !} \frac{\partial^3 f}{\partial u_1^2 \partial u_2}(\mu_1,\mu_2) (u_1-\mu_1)^2 (u_2-\mu_2)\\
 & +\frac{1}{2 !} \frac{\partial^3 f}{\partial u_1 \partial u_2^2}(\mu_1,\mu_2) (u_1-\mu_1) (u_2-\mu_2)^2
 + \frac{1}{4 !} \frac{\partial^4 f}{\partial u_1^4}(\mu_1,\mu_2) (u_1-\mu_1)^4 +\frac{1}{4 !} \frac{\partial^4 f}{\partial u_2^4}(\mu_1,\mu_2) (u_2-\mu_2)^4\\
 &+\frac{1}{3 !} \frac{\partial^4 f}{\partial u_1^3 \partial u_2}(\mu_1,\mu_2) (u_1-\mu_1)^3 (u_2-\mu_2) +\frac{1}{3 !} \frac{\partial^4 f}{\partial u_1 \partial u_2^3}(\mu_1,\mu_2) (u_1-\mu_1) (u_2-\mu_2)^3  \\
 &+ \frac{1}{2 ! 2 !} \frac{\partial^4 f}{\partial u_1^2 \partial u_2^2}(\mu_1,\mu_2) (u_1-\mu_1)^2 (u_2-\mu_2)^2  + \ldots
\end{aligned}
\end{equation}
In this case, the UDR method approximates the original function as
\begin{equation}
    \hat{f} (u_1, u_2) = f(u_1,\mu_2) + f(\mu_1,u_2) - f(\mu_1,\mu_2).
\end{equation}
The Taylor series expansion of the UDR approximation, $\hat{f}$, at  ($u_1 = \mu_1, u_2 = \mu_2$) can be expressed as
\begin{equation}
\label{eqn:2d_taylor_UDR}
    \begin{aligned}
\hat{f} (u_1, u_2) = & f(\mu_1,\mu_2)+\frac{\partial f}{\partial u_1}(\mu_1,\mu_2) (u_1- \mu_1)+\frac{\partial f}{\partial u_2}(\mu_1,\mu_2) (u_2-\mu_2) +\frac{1}{2 !} \frac{\partial^2 f}{\partial u_1^2}(\mu_1,\mu_2) (u_1-\mu_1)^2 \\
& +\frac{1}{2 !} \frac{\partial^2 f}{\partial u_2^2}(\mu_1,\mu_2) (u_2- \mu_2)^2 +\frac{1}{3 !} \frac{\partial^3 f}{\partial u_1^3}(\mu_1,\mu_2) (u_1-\mu_1)^3+\frac{1}{3 !} \frac{\partial^3 f}{\partial u_2^3}(\mu_1,\mu_2) (u_2-\mu_2)^3 \\
& +\frac{1}{4 !} \frac{\partial^4 f}{\partial u_1^4}(\mu_1,\mu_2) (u_1-\mu_1)^4+\frac{1}{4 !} \frac{\partial^4 f}{\partial u_2^4}(\mu_1,\mu_2) (u_2-\mu_2)^4+\ldots
\end{aligned}
\end{equation}
If we compare the Taylor series expansion of the original function with the UDR approximation, all of the Taylor series terms in \eqref{eqn:2d_taylor_UDR} are contained in \eqref{eqn:2d_taylor_original}, and the residual errors can be expressed as
\begin{equation}
\label{eqn: 2d_residual_error}
\begin{aligned}
    f\left(u_1, u_2\right) - \hat{f}\left(u_1, u_2\right) &=   \frac{\partial^2 f}{\partial u_1 \partial u_2}(\mu_1,\mu_2) (u_1-\mu_1) (u_2-\mu_2) +\frac{1}{2 !} \frac{\partial^3 f}{\partial u_1^2 \partial u_2}(\mu_1,\mu_2) (u_1-\mu_1)^2 (u_2-\mu_2) \\
    &
 +\frac{1}{2 !} \frac{\partial^3 f}{\partial u_1 \partial u_2^2}(\mu_1,\mu_2) (u_1-\mu_1) (u_2-\mu_2)^2
 +\frac{1}{3 !} \frac{\partial^4 f}{\partial u_1^3 \partial u_2}(\mu_1,\mu_2) (u_1-\mu_1)^3 (u_2-\mu_2) \\
 & +\frac{1}{2 ! 2 !} \frac{\partial^4 f}{\partial u_1^2 \partial u_2^2}(\mu_1,\mu_2) (u_1-\mu_1)^2 (u_2-\mu_2)^2+\frac{1}{3 !} \frac{\partial^4 f}{\partial u_1 \partial u_2^3}(\mu_1,\mu_2) (u_1-\mu_1) (u_2-\mu_2)^3 + \ldots
 \end{aligned}
\end{equation}
From \eqref{eqn: 2d_residual_error}, we observe that the UDR approximation is only second-order accurate compared to the original function. However, when it comes to estimating the mean of the output, the residual errors can be expressed as
\begin{equation}
\begin{aligned}
\label{eqn:udrerror_mean}
    \mathbb{E}[f(U_1,U_2)] -  \mathbb{E}[\hat{f}(U_1,U_2)] & =  \int_{\Gamma_1} \int_{\Gamma_2}  \left(f(u_1, u_2) - \hat{f}(u_1, u_2) \right)\rho(u_1)\rho(u_2)du_1 du_2 \\
    &= \frac{1}{2 ! 2 !} \frac{\partial^4 f}{\partial u_1^2 \partial u_2^2}(\mu_1,\mu_2)\mathbb{E}[(u_1-\mu_1)^2(u_2-\mu_2)^2] +\ldots\\
\end{aligned}
\end{equation}
In comparison, for a 3rd-order Taylor series method which approximates the function as
\begin{equation}
\label{eqn:3d_taylor}
    \begin{aligned}
\Tilde{f}\left(u_1, u_2\right)= & f(\mu_1,\mu_2)+\frac{\partial f}{\partial u_1}(\mu_1,\mu_2) (u_1-\mu_1)+\frac{\partial f}{\partial u_2}(\mu_1,\mu_2) (u_2-\mu_2) 
 +\frac{1}{2 !} \frac{\partial^2 f}{\partial u_1^2}(\mu_1,\mu_2) (u_1-\mu_1)^2 \\
 &+\frac{1}{2 !}\frac{\partial^2 f}{\partial u_2^2}(\mu_1,\mu_2) (u_2-\mu_2)^2  + \frac{\partial^2 f}{\partial u_1 \partial u_2}(\mu_1,\mu_2) (u_1-\mu_1) (u_2-\mu_2) 
 +\frac{1}{3 !} \frac{\partial^3 f}{\partial u_1^3}(\mu_1,\mu_2) (u_1-\mu_1)^3 \\
 & +\frac{1}{3 !} \frac{\partial^3 f}{\partial u_2^3}(\mu_1,\mu_2) (u_2-\mu_2)^3+\frac{1}{2 !} \frac{\partial^3 f}{\partial u_1^2 \partial u_2}(\mu_1,\mu_2) (u_1-\mu_1)^2 (u_2-\mu_2)\\
 & +\frac{1}{2 !} \frac{\partial^3 f}{\partial u_1 \partial u_2^2}(\mu_1,\mu_2) (u_1-\mu_1) (u_2-\mu_2)^2 \\
\end{aligned}
\end{equation}
The residual errors of the 3rd-order Taylor series method when estimating the mean of the output can be expressed as
\begin{equation}
\begin{aligned}
\label{eqn:t3error_mean}
    \mathbb{E}[f(U_1,U_2)] -  \mathbb{E}[\Tilde{f}(U_1,U_2)] & = \frac{1}{2 ! 2 !} \frac{\partial^4 f}{\partial u_1^2 \partial u_2^2}(\mu_1,\mu_2)\mathbb{E}[(u_1-\mu_1)^2 (u_2-\mu_2)^2] + \frac{1}{4 !} \frac{\partial^4 f}{\partial u_1^4}(\mu_1,\mu_2)\mathbb{E}[(u_1-\mu_1)^4] \\
    &+ \frac{1}{4 !} \frac{\partial^4 f}{\partial u_2^4}(\mu_1,\mu_2)\mathbb{E}[(u_2-\mu_2)^4] +\ldots
\end{aligned}
\end{equation}
Comparing the residual errors of the 3rd-order Taylor series method in \eqref{eqn:t3error_mean} with the residual errors of the UDR in \eqref{eqn:udrerror_mean}. Both of the methods are fourth-order accurate in estimating the mean of the output, but the relative errors of the UDR comprise fewer fourth-order integration terms compared with 3rd-order Taylor series expansion.
When it comes to estimating the second-order statistical moments, the relative errors of the UDR can be expressed as:
\begin{equation}
\begin{aligned}
\label{eqn:udrerror_sd}
    \mathbb{E}[f(U_1,U_2)^2] -  \mathbb{E}[\hat{f}(U_1,U_2)^2] & =  \int_{\Gamma_1} \int_{\Gamma_2}  \left(f(u_1, u_2)^2 - \hat{f}(u_1, u_2)^2 \right)\rho(u_1)\rho(u_2)du_1 du_2 \\
    &= \left(\frac{\partial^2 f}{\partial u_1 \partial u_2}(\mu_1,\mu_2)\right)^2  \mathbb{E}[(u_1-\mu_1)^2 (u_2-\mu_2)^2] \\
    &+ 
    \left(\frac{\partial y}{ \partial u_2}(\mu_1,\mu_2)
    \frac{\partial^3 y}{\partial u_1^2 \partial u_2}(\mu_1,\mu_2) +
    \frac{\partial y}{ \partial u_1}(\mu_1,\mu_2)
    \frac{\partial^3 y}{\partial u_1 \partial u_2^2}(\mu_1,\mu_2) 
    \right)\mathbb{E}[(u_1-\mu_1)^2 (u_2-\mu_2)^2]\\
 & +\frac{1}{ 2 !}f(\mu_1,\mu_2) \frac{\partial^4 f}{\partial u_1^2 \partial u_2^2}(\mu_1,\mu_2) \mathbb{E}[(u_1-\mu_1)^2 (u_2-\mu_2)^2] +\ldots
\end{aligned}
\end{equation}
The residual errors for 3rd order Taylor series method can be expressed as:
\begin{equation}
\label{eqn:t3error_sd}
     \mathbb{E}[f(U_1,U_2)^2] -  \mathbb{E}[\Tilde{f}(U_1,U_2)^2] = \frac{1}{ 2 !}f(\mu_1,\mu_2) \frac{\partial^4 f}{\partial u_1^2 \partial u_2^2}(\mu_1,\mu_2) \mathbb{E}[(u_1-\mu_1)^2 (u_2-\mu_2)^2] +\ldots
\end{equation}
Comparing the residual errors of the 3rd-order Taylor series expansion method in \eqref{eqn:t3error_sd} with the residual errors of the UDR in \eqref{eqn:udrerror_sd}. Both of the methods are also fourth-order accurate in estimating the second-order moment of the output, but the relative errors of the UDR comprise more fourth-order integration terms compared with 3rd-order Taylor series expansion.

\subsection{Gradient-enhanced univariate dimension reduction in 2D case}
\label{Sec: GUDR_2D}
In this section, we derive the residual errors of the gradient-enhanced univariate dimension reduction (GUDR) method in estimating the mean and standard deviation for a 2D UQ problem.  
Consider an UQ problem that involves a bivariate function $f\left(u_1, u_2\right)$, the GUDR method approximates the original function as
\begin{equation}
\begin{aligned}
    \Bar{f}(u_1,u_2) &= y(u_1,\mu_2) + y(\mu_1,u_2) - y(\mu_1,\mu_2) + (u_1-\mu_1) \left(\frac{\partial f}{\partial u_1}(\mu_1,u_2) - \frac{\partial f}{\partial u_1}(\mu_1,\mu_2) \right)  \\
    & + (u_2-\mu_2)\left(\frac{\partial f}{\partial u_2}(u_1,\mu_2)  - \frac{\partial f}{\partial u_2}(\mu_1,\mu_2)\right) -  \frac{\partial^2 f}{\partial u_1 u_2}(\mu_1,\mu_2)(u_1-\mu_1)(u_2-\mu_2). 
\end{aligned}
\end{equation}
The Taylor series of the involved univariate gradient terms at ($u_1 = \mu_1, u_2 = \mu_2$) can be expressed by
\begin{equation}
    \frac{\partial f}{\partial u_1}(\mu_1,u_2) = \frac{\partial f}{\partial u_1}(\mu_1,\mu_2)  + \frac{\partial^2 f}{\partial u_1 u_2}(\mu_1,\mu_2)(u_2-\mu_2) +  \frac{1}{2 !}\frac{\partial^3 f}{\partial u_1 u_2^2}(\mu_1,\mu_2)(u_2-\mu_2)^2+ \frac{1}{3 !}\frac{\partial^4 f}{\partial u_1 u_2^3}(\mu_1,\mu_2)(u_2-\mu_2)^3 + \ldots,
\end{equation}
\begin{equation}
    \frac{\partial f}{\partial u_2}(u_1,\mu_2) = \frac{\partial f}{\partial u_2}(\mu_1,\mu_2)  + \frac{\partial^2 f}{\partial u_1 u_2}(\mu_1,\mu_2)(u_1-\mu_1) +  \frac{1}{2 !}\frac{\partial^3 f}{\partial u_1^2 u_2}(\mu_1,\mu_2)(u_1-\mu_1)^2+ \frac{1}{3 !}\frac{\partial^4 f}{\partial u_1^3 u_2}(\mu_1,\mu_2)(u_1-\mu_1)^3 + \ldots.
\end{equation}
The Taylor series of GUDR approximation function, $\Bar{f}$ can be expressed as
\begin{equation}
\label{eqn:2d_taylor_GUDR}
    \begin{aligned}
\Bar{f}\left(u_1, u_2\right)= & f(\mu_1,\mu_2)+\frac{\partial f}{\partial u_1}(\mu_1,\mu_2) (u_1-\mu_1)+\frac{\partial f}{\partial u_2}(\mu_1,\mu_2) (u_2-\mu_2) 
 +\frac{1}{2 !} \frac{\partial^2 f}{\partial u_1^2}(\mu_1,\mu_2) (u_1-\mu_1)^2 \\
 &+\frac{1}{2 !}\frac{\partial^2 f}{\partial u_2^2}(\mu_1,\mu_2) (u_2-\mu_2)^2  + \frac{\partial^2 f}{\partial u_1 \partial u_2}(\mu_1,\mu_2) (u_1-\mu_1) (u_2-\mu_2) 
 +\frac{1}{3 !} \frac{\partial^3 f}{\partial u_1^3}(\mu_1,\mu_2) (u_1-\mu_1)^3 \\
 & +\frac{1}{3 !} \frac{\partial^3 f}{\partial u_2^3}(\mu_1,\mu_2) (u_2-\mu_2)^3+\frac{1}{2 !} \frac{\partial^3 f}{\partial u_1^2 \partial u_2}(\mu_1,\mu_2) (u_1-\mu_1)^2 (u_2-\mu_2)\\
 & +\frac{1}{2 !} \frac{\partial^3 f}{\partial u_1 \partial u_2^2}(\mu_1,\mu_2) (u_1-\mu_1) (u_2-\mu_2)^2
 + \frac{1}{4 !} \frac{\partial^4 f}{\partial u_1^4}(\mu_1,\mu_2) (u_1-\mu_1)^4 +\frac{1}{4 !} \frac{\partial^4 f}{\partial u_2^4}(\mu_1,\mu_2) (u_2-\mu_2)^4\\
 &+\frac{1}{3 !} \frac{\partial^4 f}{\partial u_1^3 \partial u_2}(\mu_1,\mu_2) (u_1-\mu_1)^3 (u_2-\mu_2) +\frac{1}{3 !} \frac{\partial^4 f}{\partial u_1 \partial u_2^3}(\mu_1,\mu_2) (u_1-\mu_1) (u_2-\mu_2)^3  + \ldots
\end{aligned}
\end{equation}
If we compare the Taylor series expansion of the original function with the GUDR approximation, all of the Taylor series terms in \eqref{eqn:2d_taylor_GUDR} are contained in \eqref{eqn:2d_taylor_original}, and the residual errors can be expressed as
 \begin{equation}
\begin{aligned}
    f\left(u_1, u_2\right) - \bar{f}\left(u_1, u_2\right) &=   \frac{1}{2 ! 2 !} \frac{\partial^4 f}{\partial u_1^2 \partial u_2^2}(0,0) u_1^2 u_2^2 +\dots
 \end{aligned}
\end{equation}
which only include fourth and higher-order terms.
When we estimate the first-order statistical moment of the output, the GUDR approximation generates the same results as the UDR as
\begin{equation}
\label{eqn:gudrerror_mean}
    \mathbb{E}[f(u_1,u_2)] -  \mathbb{E}[\bar{f}(u_1,u_2)] 
    = \frac{1}{2 ! 2 !} \frac{\partial^4 f}{\partial u_1^2 \partial u_2^2}(\mu_1,\mu_2)\mathbb{E}[(u_1-\mu_1)^2(u_2-\mu_2)^2] +\ldots\\
\end{equation}
When we estimate the second-order statistical moment of the output, the residual errors for GUDR approximation can be expressed as
\begin{equation}
\label{eqn:t3error_sd}
     \mathbb{E}[f(u_1,u_2)^2] -  \mathbb{E}[\Bar{f}(u_1,u_2)^2] = \frac{1}{ 2 !}f(\mu_1,\mu_2) \frac{\partial^4 f}{\partial u_1^2 \partial u_2^2}(\mu_1,\mu_2) \mathbb{E}[(u_1-\mu_1)^2 (u_2-\mu_2)^2] +\ldots.
\end{equation}
Compared to the residual errors of the 3rd-order Taylor series method, shown in \eqref{eqn:t3error_sd}, the residual errors of the GUDR approximation contain the same fourth-order integration terms.

\subsection{Example code to implement computational graph transformation in GUDR}
\label{sec: python_code}
In this section, we provide the example code to implement the computational graph transformation method in GUDR for a 2D UQ problem.
The Python code below assumes we have the univariate model and gradient functions' evaluations on quadrature points of each input dimension, along with the function and gradient evaluations on the mean of the input.
The code shows how to use NumPy's Einstein summation function to expand the sizes of certain inputs so that we can generate the correct tensor-grid evaluations for the GUDR approximation function.

\begin{python}
import numpy as np
k # number of quadrature points in each dimension
u_1 # quadrature points in u_1, size: (k,1)
u_2  # quadrature points in u_2, size: (k,1)
f1_u1 #  quadrature points evaluations of f_1(u_1), size (k,1)
f2_u2 #  quadrature points evaluations of f_2(u_2), size (k,1)
df1_du # quadrature points evaluations of df_1/du(u_1), size (k,2)
df2_du # quadrature points evaluations of df_2/du(u_2), size (k,2)
mean_u # mean of the inputs, size(2,)
f_mean_u = # function evaluation on mean of the inputs, size (1,)
df_du_mean_u = # gradient evaluation on mean of the inputs, size (2,)
d2f_du2_mean_u = # Hessian evaluation on mean of the inputs, size (2,2)
total_points = k**2 # total input points k**2

# Use Einsum and reshape operations to expand the size of certain inputs
u_1 = np.einsum('i...,p...->ip...', u_1 , np.ones(k))
u_1 = np.reshape(u_1 , (total_points,))
u_2 = np.einsum('p...,i...->ip...', u_2 , np.ones(k))
u_2 = np.reshape(u_2 , (total_points,))
f1_u1 = np.einsum('i...,p...->ip...', f1_u1, np.ones(k))
f1_u1 = np.reshape(f1_u1, (total_points,))
f2_u2 = np.einsum('p...,i...->ip...', f2_u2, np.ones(k))
f2_u2 = np.reshape(f2_u2, (total_points,))
df1_du = np.einsum('i...,p...->ip...', df1_du, np.ones(k))
df1_du = np.reshape(df1_du, (total_points, 2))
df2_du = np.einsum('p...,i...->ip...', df2_du, np.ones(k))
df2_du = np.reshape(df2_du, (total_points, 2))

# Evaluate GUDR approximation function on tensor-grid quadrature points
f_gudr = np.zeros(total_points,) 
for i in range(total_points):
  u =  np.array([u_1[i], u_2[i]])
  a1 = np.array([0, u_2[i]- mean_u[1]])
  a2 = np.array([u_1[i]- mean_u[0], 0])
  term1 = f1_u1[i] + f2_u2[i] - f_mean_u
  term2 = np.dot(a1, df1_du[i,:]) + np.dot(a2, df2_du[i,:])- np.dot((u-mean_u), df_du_mean_u)
  term3 = -0.5*np.einsum('...i,...i->...', (u-mean_u).T.dot(d2f_du2_mean_u - np.diag(np.diag(d2f_du2_mean_u))), (u-mean_u))
  f_gudr[i] = term1 + term2+ term3

# f_gudr: GUDR approximation function evaluations on tensor-grid quadrature points, size (k^2,)
\end{python}
\section*{Acknowledgments}

The material presented in this paper is, in part, based upon work supported by  DARPA under grant No.~D23AP00028-00.

\section*{Declaration of interests}
\urlstyle{rm}
This paper was originally presented as Paper 2024-1234 at the AIAA SCITECH 2024 Forum, Orlando FL, January 8-12. 2024. Copyright by Bingran Wang. Published by the American Institute of Aeronautics and Astronautics, Inc., with permission.  doi: \url{https://doi.org/10.2514/6.2024-1234} 
\bibliography{sample}

\begin{thebibliography}{38}
\newcommand{\enquote}[1]{``#1''}
\providecommand{\natexlab}[1]{#1}
\providecommand{\url}[1]{\texttt{#1}}
\providecommand{\urlprefix}{URL }
\expandafter\ifx\csname urlstyle\endcsname\relax
  \providecommand{\doi}[1]{\discretionary{}{}{}https://doi.org/#1}\else
  \providecommand{\doi}[1]{\discretionary{}{}{}\urlstyle{rm}\url{https://doi.org/#1}}\fi

\bibitem[{Joslyn and Savelli(2010)}]{joslyn2010communicating}
Joslyn, S., and Savelli, S., \enquote{Communicating forecast uncertainty: Public perception of weather forecast uncertainty,} \emph{Meteorological Applications}, Vol.~17, No.~2, 2010, pp. 180--195.
\newblock \doi{10.1002/met.190}.

\bibitem[{Pappenberger et~al.(2005)Pappenberger, Beven, Hunter, Bates, Gouweleeuw, Thielen, and de~Roo}]{hess-9-381-2005}
Pappenberger, F., Beven, K.~J., Hunter, N.~M., Bates, P.~D., Gouweleeuw, B.~T., Thielen, J., and de~Roo, A. P.~J., \enquote{Cascading model uncertainty from medium range weather forecasts (10 days) through a rainfall-runoff model to flood inundation predictions within the European Flood Forecasting System (EFFS),} \emph{Hydrology and Earth System Sciences}, Vol.~9, No.~4, 2005, pp. 381--393.
\newblock \doi{10.5194/hess-9-381-2005}.

\bibitem[{H{\"u}llermeier and Waegeman(2021)}]{hullermeier2021aleatoric}
H{\"u}llermeier, E., and Waegeman, W., \enquote{Aleatoric and epistemic uncertainty in machine learning: An introduction to concepts and methods,} \emph{Machine Learning}, Vol. 110, 2021, pp. 457--506.
\newblock \doi{10.1007/s10994-021-05946-3}.

\bibitem[{Wan et~al.(2014)Wan, Mao, Todd, and Ren}]{wan2014analytical}
Wan, H.-P., Mao, Z., Todd, M.~D., and Ren, W.-X., \enquote{Analytical uncertainty quantification for modal frequencies with structural parameter uncertainty using a Gaussian process metamodel,} \emph{Engineering Structures}, Vol.~75, 2014, pp. 577--589.
\newblock \doi{10.1016/j.engstruct.2014.06.028}.

\bibitem[{Hu et~al.(2018)Hu, Mahadevan, and Ao}]{hu2018uncertainty}
Hu, Z., Mahadevan, S., and Ao, D., \enquote{Uncertainty aggregation and reduction in structure--material performance prediction,} \emph{Computational Mechanics}, Vol.~61, No.~1, 2018, pp. 237--257.
\newblock \doi{10.1007/s00466-017-1448-6}.

\bibitem[{Ng and Willcox(2016)}]{ng2016monte}
Ng, L.~W., and Willcox, K.~E., \enquote{Monte Carlo information-reuse approach to aircraft conceptual design optimization under uncertainty,} \emph{Journal of Aircraft}, Vol.~53, No.~2, 2016, pp. 427--438.
\newblock \doi{10.2514/1.C033352}.

\bibitem[{Wang et~al.(2024{\natexlab{a}})Wang, Orndorff, Joshy, and Hwang}]{wang2024graph}
Wang, B., Orndorff, N.~C., Joshy, A.~J., and Hwang, J.~T., \enquote{Graph-accelerated large-scale multidisciplinary design optimization under uncertainty of a laser-beam-powered aircraft,} \emph{AIAA SCITECH 2024 Forum}, 2024{\natexlab{a}}, p. 0169.
\newblock \doi{10.2514/6.2024-0169}.

\bibitem[{Lim et~al.(2022)Lim, Kim, and Yee}]{lim2022uncertainty}
Lim, D., Kim, H., and Yee, K., \enquote{Uncertainty propagation in flight performance of multirotor with parametric and model uncertainties,} \emph{Aerospace Science and Technology}, Vol. 122, 2022, p. 107398.
\newblock \doi{10.1016/j.ast.2022.107398}.

\bibitem[{Wooldridge(2001)}]{wooldridge2001applications}
Wooldridge, J.~M., \enquote{Applications of generalized method of moments estimation,} \emph{Journal of Economic perspectives}, Vol.~15, No.~4, 2001, pp. 87--100.
\newblock \doi{10.1257/jep.15.4.87}.

\bibitem[{Fragkos et~al.(2019)Fragkos, Papoutsis-Kiachagias, and Giannakoglou}]{fragkos2019pfosm}
Fragkos, K., Papoutsis-Kiachagias, E., and Giannakoglou, K., \enquote{pFOSM: An efficient algorithm for aerodynamic robust design based on continuous adjoint and matrix-vector products,} \emph{Computers \& Fluids}, Vol. 181, 2019, pp. 57--66.
\newblock \doi{10.1016/j.compfluid.2019.01.016}.

\bibitem[{Peherstorfer et~al.(2016)Peherstorfer, Willcox, and Gunzburger}]{peherstorfer2016optimal}
Peherstorfer, B., Willcox, K., and Gunzburger, M., \enquote{Optimal model management for multifidelity Monte Carlo estimation,} \emph{SIAM Journal on Scientific Computing}, Vol.~38, No.~5, 2016, pp. A3163--A3194.
\newblock \doi{10.1137/15M1046472}.

\bibitem[{Peherstorfer et~al.(2018)Peherstorfer, Willcox, and Gunzburger}]{peherstorfer2018survey}
Peherstorfer, B., Willcox, K., and Gunzburger, M., \enquote{Survey of multifidelity methods in uncertainty propagation, inference, and optimization,} \emph{Siam Review}, Vol.~60, No.~3, 2018, pp. 550--591.
\newblock \doi{10.1137/16M1082469}.

\bibitem[{Tabandeh et~al.(2022)Tabandeh, Jia, and Gardoni}]{tabandeh2022review}
Tabandeh, A., Jia, G., and Gardoni, P., \enquote{A review and assessment of importance sampling methods for reliability analysis,} \emph{Structural Safety}, Vol.~97, 2022, p. 102216.
\newblock \doi{10.1016/j.strusafe.2022.102216}.

\bibitem[{Kaymaz(2005)}]{kaymaz2005application}
Kaymaz, I., \enquote{Application of kriging method to structural reliability problems,} \emph{Structural safety}, Vol.~27, No.~2, 2005, pp. 133--151.
\newblock \doi{10.1016/j.strusafe.2004.09.001}.

\bibitem[{Hu and Mahadevan(2016)}]{hu2016single}
Hu, Z., and Mahadevan, S., \enquote{A single-loop kriging surrogate modeling for time-dependent reliability analysis,} \emph{Journal of Mechanical Design}, Vol. 138, No.~6, 2016, p. 061406.
\newblock \doi{10.1115/1.4033428}.

\bibitem[{Rumpfkeil(2013)}]{rumpfkeil2013optimizations}
Rumpfkeil, M.~P., \enquote{Optimizations under uncertainty using gradients, Hessians, and surrogate models,} \emph{AIAA journal}, Vol.~51, No.~2, 2013, pp. 444--451.
\newblock \doi{10.2514/1.J051847}.

\bibitem[{Hosder et~al.(2006)Hosder, Walters, and Perez}]{hosder2006non}
Hosder, S., Walters, R., and Perez, R., \enquote{A non-intrusive polynomial chaos method for uncertainty propagation in CFD simulations,} \emph{44th AIAA aerospace sciences meeting and exhibit}, 2006, p. 891.
\newblock \doi{10.2514/6.2006-891}.

\bibitem[{Jones et~al.(2013)Jones, Doostan, and Born}]{jones2013nonlinear}
Jones, B.~A., Doostan, A., and Born, G.~H., \enquote{Nonlinear propagation of orbit uncertainty using non-intrusive polynomial chaos,} \emph{Journal of Guidance, Control, and Dynamics}, Vol.~36, No.~2, 2013, pp. 430--444.
\newblock \doi{10.2514/1.57599}.

\bibitem[{Keshavarzzadeh et~al.(2017)Keshavarzzadeh, Fernandez, and Tortorelli}]{keshavarzzadeh2017topology}
Keshavarzzadeh, V., Fernandez, F., and Tortorelli, D.~A., \enquote{Topology optimization under uncertainty via non-intrusive polynomial chaos expansion,} \emph{Computer Methods in Applied Mechanics and Engineering}, Vol. 318, 2017, pp. 120--147.
\newblock \doi{10.1016/j.cma.2017.01.019}.

\bibitem[{Morio(2011)}]{morio2011global}
Morio, J., \enquote{Global and local sensitivity analysis methods for a physical system,} \emph{European journal of physics}, Vol.~32, No.~6, 2011, p. 1577.
\newblock \doi{10.1088/0143-0807/32/6/011}.

\bibitem[{Chun and Kele{\c{s}}(2010)}]{chun2010sparse}
Chun, H., and Kele{\c{s}}, S., \enquote{Sparse partial least squares regression for simultaneous dimension reduction and variable selection,} \emph{Journal of the Royal Statistical Society: Series B (Statistical Methodology)}, Vol.~72, No.~1, 2010, pp. 3--25.
\newblock \doi{10.1111/j.1467-9868.2009.00723.x}.

\bibitem[{Constantine et~al.(2014)Constantine, Dow, and Wang}]{constantine2014active}
Constantine, P.~G., Dow, E., and Wang, Q., \enquote{Active subspace methods in theory and practice: applications to kriging surfaces,} \emph{SIAM Journal on Scientific Computing}, Vol.~36, No.~4, 2014, pp. A1500--A1524.
\newblock \doi{10.1137/130916138}.

\bibitem[{Rahman and Xu(2004)}]{rahman2004univariate}
Rahman, S., and Xu, H., \enquote{A univariate dimension-reduction method for multi-dimensional integration in stochastic mechanics,} \emph{Probabilistic Engineering Mechanics}, Vol.~19, No.~4, 2004, pp. 393--408.
\newblock \doi{10.1016/j.probengmech.2004.04.003}.

\bibitem[{Piric(2015)}]{piric2015reliability}
Piric, K., \enquote{Reliability analysis method based on determination of the performance function’s PDF using the univariate dimension reduction method,} \emph{Structural safety}, Vol.~57, 2015, pp. 18--25.
\newblock \doi{10.1016/j.strusafe.2015.07.005}.

\bibitem[{Huang and Du(2006)}]{huang2006uncertainty}
Huang, B., and Du, X., \enquote{Uncertainty analysis by dimension reduction integration and saddlepoint approximations,} 2006.
\newblock \doi{10.1115/1.2118667}.

\bibitem[{Lee et~al.(2008)Lee, Choi, Du, and Gorsich}]{lee2008dimension}
Lee, I., Choi, K., Du, L., and Gorsich, D., \enquote{Dimension reduction method for reliability-based robust design optimization,} \emph{Computers \& Structures}, Vol.~86, No. 13-14, 2008, pp. 1550--1562.
\newblock \doi{10.1016/j.compstruc.2007.05.020}.

\bibitem[{Xu and Rahman(2004)}]{xu2004generalized}
Xu, H., and Rahman, S., \enquote{A generalized dimension-reduction method for multidimensional integration in stochastic mechanics,} \emph{International Journal for Numerical Methods in Engineering}, Vol.~61, No.~12, 2004, pp. 1992--2019.
\newblock \doi{10.1002/nme.1135}.

\bibitem[{Wang et~al.(2024{\natexlab{b}})Wang, Sperry, Gandarillas, and Hwang}]{wang2023accelerating}
Wang, B., Sperry, M., Gandarillas, V.~E., and Hwang, J.~T., \enquote{Accelerating model evaluations in uncertainty propagation on tensor grids using computational graph transformations,} \emph{Aerospace Science and Technology}, Vol. 145, 2024{\natexlab{b}}, p. 108843.
\newblock \doi{10.1016/j.ast.2023.108843}.

\bibitem[{Wang et~al.(2022)Wang, Sperry, Gandarillas, and Hwang}]{wang2022efficient}
Wang, B., Sperry, M., Gandarillas, V.~E., and Hwang, J.~T., \enquote{Efficient uncertainty propagation through computational graph modification and automatic code generation,} \emph{AIAA AVIATION 2022 Forum}, 2022, p. 3997.
\newblock \doi{10.2514/6.2022-3997}.

\bibitem[{Wang et~al.(2024{\natexlab{c}})Wang, Orndorff, Sperry, and Hwang}]{wang2024extension}
Wang, B., Orndorff, N.~C., Sperry, M., and Hwang, J.~T., \enquote{Extension of graph-accelerated non-intrusive polynomial chaos to high-dimensional uncertainty quantification through the active subspace method,} \emph{arXiv preprint arXiv:2405.05556}, 2024{\natexlab{c}}.
\newblock \doi{10.48550/arXiv.2405.05556}.

\bibitem[{Wang et~al.(2024{\natexlab{d}})Wang, Orndorff, and Hwang}]{wang2024graphpartial}
Wang, B., Orndorff, N.~C., and Hwang, J.~T., \enquote{Graph-accelerated non-intrusive polynomial chaos expansion using partially tensor-structured quadrature rules,} \emph{arXiv preprint arXiv:2403.15614}, 2024{\natexlab{d}}.
\newblock \doi{10.48550/arXiv.2403.15614}.

\bibitem[{Wang(2024)}]{wang2024graphthesis}
Wang, B., \enquote{Graph-accelerated uncertainty propagation for large-scale multidisciplinary design, analysis, and optimization under uncertainty,} Ph.D. thesis, UC San Diego, 2024.

\bibitem[{Gandarillas et~al.(2024)Gandarillas, Joshy, Sperry, Ivanov, and Hwang}]{gandarillas2022novel}
Gandarillas, V., Joshy, A.~J., Sperry, M.~Z., Ivanov, A.~K., and Hwang, J.~T., \enquote{A graph-based methodology for constructing computational models that automates adjoint-based sensitivity analysis,} \emph{Structural and Multidisciplinary Optimization}, Vol.~67, No.~5, 2024, p.~76.
\newblock \doi{10.1007/s00158-024-03792-0}.

\bibitem[{Christianson(1992)}]{christianson1992automatic}
Christianson, B., \enquote{Automatic Hessians by reverse accumulation,} \emph{IMA Journal of Numerical Analysis}, Vol.~12, No.~2, 1992, pp. 135--150.
\newblock \doi{10.1093/imanum/12.2.135}.

\bibitem[{Ruh and Hwang(2023)}]{ruh2023fast}
Ruh, M.~L., and Hwang, J.~T., \enquote{Fast and Robust Computation of Optimal Rotor Designs Using Blade Element Momentum Theory,} \emph{AIAA Journal}, Vol.~61, No.~9, 2023, pp. 4096--4111.
\newblock \doi{10.2514/1.J062611}.

\bibitem[{Sperry et~al.(2023)Sperry, Kondap, and Hwang}]{sperry2023automatic}
Sperry, M., Kondap, K., and Hwang, J.~T., \enquote{Automatic adjoint sensitivity analysis of models for large-scale multidisciplinary design optimization,} \emph{AIAA AVIATION 2023 Forum}, 2023, p. 3721.
\newblock \doi{10.2514/6.2023-3721}.

\bibitem[{Feinberg and Langtangen(2015)}]{feinberg2015chaospy}
Feinberg, J., and Langtangen, H.~P., \enquote{Chaospy: An open source tool for designing methods of uncertainty quantification,} \emph{Journal of Computational Science}, Vol.~11, 2015, pp. 46--57.
\newblock \doi{10.1016/j.jocs.2015.08.008}.

\bibitem[{Orndorff et~al.(2023)Orndorff, Wang, Ruh, Fletcher, and Hwang}]{orndorff2023gradient}
Orndorff, N.~C., Wang, B., Ruh, M.~L., Fletcher, A., and Hwang, J.~T., \enquote{Gradient-based sizing optimization of power-beaming-enabled aircraft,} \emph{AIAA AVIATION 2023 Forum}, 2023, p. 4019.
\newblock \doi{10.2514/6.2023-4019}.

\end{thebibliography}
\end{document}